\documentclass[
    aps,
    prb,
    reprint,
    onecolumn,
    10pt,
    superscriptaddress,
    nofootinbib
]{revtex4-2}

\usepackage{amsmath,amssymb}
\usepackage{graphicx}
\usepackage{bm}
\usepackage[hidelinks]{hyperref}
\usepackage{geometry}
\usepackage{float} 
\usepackage{microtype}

\numberwithin{equation}{section}
\renewcommand{\theequation}{\arabic{section}.\arabic{equation}}

\begin{document}

\title{Analytic and Numerical Investigation of a Proposed Non-linear Extension of Quantum Mechanics}

\author{Alan Chodos}
\email{alan.chodos@uta.edu}

\author{Anastasiia Purtova}
\affiliation{Department of Physics, University of Texas at Arlington, 
502 Yates Street, Box 19059, Arlington, Texas 76019}

\begin{abstract}
In this paper, additional properties of a previously introduced non-linear extension of quantum mechanics are investigated. Analytically, we uncover a limit in which the model reduces to ordinary linear quantum mechanics. We also perform a perturbation analysis about an exact solution studied in earlier work. The equations represent a system that is subject to a periodic driver, and we use Floquet theory to help determine its properties. Numerically, we compare the exact solution to the analytic approximation, and we exhibit several examples of how the non-linear effects influence particle trajectories, with particular emphasis on trajectories near the linear limit.
\end{abstract}

%%%%%%%%%%%%%%%%%%%%%%%%%%%%%%%%%%%%%%%%%%
\date{\today}
\maketitle
%%%%%%%%%%%%%%%%%%%%%%%%%%%%%%%%%%%%%%%%%%
\section{Introduction}
The Schrödinger equation of ordinary quantum mechanics is linear in the state vector, a property that gives rise to the superposition principle. From time to time, attempts have been made to construct non-linear generalizations, going back at least to Kibble in the 1970s \cite{Kibble1978,Kibble1979}. Kibble’s idea was to replace certain coupling constants in field theory with expectation values of operators, thereby introducing state-dependence into the Lagrangian. A renaissance of this approach has emerged recently in the work of Kaplan and Rajendran \cite{KaplanRajendran2022, KaplanRajendran2025}.

A well-known non-linear generalization was that of Weinberg \cite{Weinberg1989}, whose motivation was to test quantum mechanics by deriving predictions from his model that could be probed experimentally. However, as pointed out by Gisin \cite{Gisin1990} and Polchinski \cite{Polchinski1991}, Weinberg’s model allowed superluminal communication, and as Polchinski also noticed, in the context of the many-worlds interpretation, it allowed communication between different branches of the wave function. As a result, Weinberg’s model languished, and, as far as we are aware, the experimental tests he proposed were never done \cite{Hsu2026}. We remark, however, that despite widespread theoretical prejudice, superluminal communication, like anything else, needs to be tested experimentally. A model should not be ruled out a priori unless its predictions contradict already established experimental bounds.

One motivation for making quantum mechanics non-linear is the possible relevance of such an extension to the relationship between quantum mechanics and gravity, a famously difficult problem that has not yet attained a generally accepted resolution. Although most attempts, such as string theory and loop quantum gravity, have sought to incorporate general relativity within the confines of ordinary quantum mechanics, others \cite{Penrose2014, Carlip2008} have argued that a successful marriage might require a modification of the rules of quantum mechanics as well. Here we begin to investigate the possibility that some, or possibly all, the properties of gravity can emerge from a suitably chosen quantum mechanical extension, without the manifest use of the machinery of general relativity. Our strategy is to study how the non-linear dynamics influences particle trajectories, expressed as the expectation values of position operators, and see whether there is any resemblance to the motion of a particle in a gravitational field. 

The non-linear extension of quantum mechanics we shall use has been introduced in previous work \cite{ChodosCooper2023, ChodosCooper2024, ChodosCooper2025}. The scheme departs fundamentally from ordinary quantum mechanics by employing a pair of state vectors to describe a given system, rather than the single state vector conventionally used for a pure state. Thus, the system inherently involves a mixture of two states. However, the components do not evolve independently under the action of some Hamiltonian. Rather, the equations of motion couple the two states $|\psi_1 \rangle$  and $|\psi_2 \rangle$ together via the inner products $\langle \psi_i|\psi_j \rangle, i,j=1,2,$ thereby giving rise to the non-linearity. 

In this work we present new analytical and numerical results. On the analytic side, we derive the limit in which our system reduces to ordinary quantum mechanics. We also find the solution in a special case that was left untreated in previous work. And we derive a perturbation expansion about an exact solution studied in reference \cite{ChodosCooper2025}. We also use the periodicity built into the equations to perform a Floquet analysis of the solutions.
Numerically, we give representative solutions for the wave functions and derive therefrom samples of the trajectories alluded to above. We also compare the analytical approximations to the numerical integration of the exact equations.

In section \ref{sec:model_review} we review the essence of our model. In section \ref{sec:analytic_res}, we discuss the linear limit, solve the exceptional case, and derive the perturbation expansion. In section \ref{sec:floquet_structure} we use Floquet theory to analyze our equations. In section \ref{sec:numeric_res} we present and discuss our numerical results. Section \ref{sec:conclusions} is devoted to some conclusions.

\section{Review of The Model}
\label{sec:model_review}
We begin with a concise review of the non-linear extension of quantum mechanics that was studied in reference \cite{ChodosCooper2025}. We consider a Hilbert space on which is defined a Hamiltonian operator $H$. We can accommodate any system of interest, from a single particle to a relativistic quantum field theory, to something else, provided it has a Hamiltonian description. But instead of introducing a single state vector that satisfies the usual Schrödinger equation, we introduce a pair of state vectors, $| \psi_i \rangle, i=1,2,$ and postulate the following equations of motion:
\begin{equation}
\begin{aligned}
    i \frac{d|\psi_1\rangle}{dt} &= \left[ H+ i \mu \langle \psi_2 | \psi_2 \rangle\right] |\psi_1 \rangle + (g+\lambda) \langle \psi_2| \psi_1 \rangle |\psi_2 \rangle \\
    i \frac{d|\psi_2\rangle}{dt} &= \left[ H - i \mu \langle \psi_1 | \psi_1 \rangle\right] |\psi_2 \rangle + (g^*-\lambda) \langle \psi_1| \psi_2 \rangle |\psi_1 \rangle 
    \end{aligned}
    \label{eq:general_system}
\end{equation}
Here $\mu$ and $\lambda$ are real parameters, and $g=a+ i b$ is complex. These equations are consistent with the following 3 principles:
\begin{enumerate}
    \item $\frac{d}{dt} \left[ \langle \psi_1|\psi_1\rangle + \langle \psi_2 | \psi_2 \rangle \right]=0.$
    \item The extra terms should be linear in the inner products $\langle \psi_i|\psi_j\rangle, i,j=1,2,$ and depend on time only through them. 
    \item The system should possess the symmetries $|\psi_j\rangle \rightarrow e^{i \theta_j} |\psi_j \rangle,~ j=1,2$ for constant $\theta_j$. 
\end{enumerate}
Principle 1. is the generalization to our system of the requirement that the norm of the state vector is conserved. It is important in guaranteeing that the solutions to the equations of motion remain bounded. Principle 2. distinguishes our strategy from other possible ways of rendering quantum mechanics non-linear. Principle 3. restricts the number of terms that can appear in the equations of motion. Even so, our equations are not quite the most general. As described in reference \cite{ChodosCooper2025}, there are 4 additional terms that could have been included, but these seem to have little influence over the dynamics, and in this work for simplicity we set them to zero.

Setting $\langle \psi_1|\psi_1\rangle+\langle \psi_2|\psi_2\rangle = N$, we define the density matrix 
\begin{equation}
    \hat \rho (t) = \frac{1}{N} \left[ |\psi_1(t) \rangle\langle \psi_1(t) | + |\psi_2 (t) \rangle \langle \psi_2 (t)| \right]
    \label{eq:density_matrix_operator}
\end{equation}
and we postulate that the expectation value of any operator $X$ is given by
\begin{equation}
    \langle X \rangle (t) = \operatorname{Tr} \hat\rho (t) X.
    \label{eq:X_expectation}
\end{equation}
Together, the equations of motion and the formula for the expectation value of an operator define the model. Everything that follows is a consequence of these postulates.\footnote{Of course, as in ordinary quantum mechanics, further discussion, and possibly an additional postulate is needed, to describe what happens as a result of measurement. Do our state vectors “collapse”, and, if so, how? This important question is deferred to future work.}

In this paper, we shall calculate $\langle X\rangle (t)$ for some simple instructive examples. To focus on the effects of the non-linearity, we consider the case where the underlying system is very simple, a single free particle, and where $X_k$ is the position operator for that particle. 

We shall find below that for appropriate choices of the input parameters, $\langle X_k \rangle (t)$ describes bounded, quasi-periodic orbits, whose properties we examine. 

The route from the equations of motion to $\langle X_k \rangle (t)$ involves a series of 3 steps. As in previous work, we define 
\begin{equation}
    \tau = \langle \psi_1|\psi_1\rangle - \langle \psi_2|\psi_2\rangle, \quad \gamma = \langle \psi_2|\psi_1\rangle
    \label{eq:tau_and_gamma_def}
\end{equation}
and
\begin{equation}
    \delta = | \gamma|^2.
    \label{eq:delta_def}
\end{equation}
In step 1, from the equations of motion we deduce a pair of first-order non-linear equations for the $\tau-\delta$ system, independent of the dynamics of the underlying system. We show that the $\tau-\delta$ equations can be cast in Hamiltonian form, with $\kappa=\ln \delta$ the coordinate and $\tau$ the momentum, and a conserved energy given by $N^2$. This is an independent subsystem, acting like an internal driver that dictates the behavior of the full system of equations. The $\tau-\delta$ equations also allow us to solve for $\gamma$. For our choice of parameters, the driver is a periodic system.

In step 2, we insert the information from the $\tau-\delta$  driver into the full equations of motion. These now take the form of a pair of linear coupled first-order equations for the $|\psi_i (t) \rangle$, which will have 2 linearly independent solutions. To accommodate them, we introduce two vectors $|A_i (t) \rangle$ that evolve according to the underlying Hamiltonian (in this case just a free particle), normalized as $\langle A_i|A_j \rangle=\delta_{ij}$, and express the solutions for the $|\psi_i (t) \rangle$ as linear combinations of the $|A_j (t) \rangle$ with time-dependent coefficients that are determined by the equations of motion. It is these coefficients that will be crucial in evaluating $\langle X_i\rangle (t)$. The coefficients depend on overall constants of integration, which we fix by imposing the normalization conditions
\begin{equation}
    \langle \psi_1|\psi_1 \rangle + \langle \psi_2|\psi_2 \rangle = N, \quad \langle \psi_1|\psi_1 \rangle - \langle \psi_2|\psi_2 \rangle = \tau,  \quad\text{ and }\quad \langle \psi_2|\psi_1 \rangle = \gamma.
    \label{eq:normalization_condtions}
\end{equation}
Step 3 is the evaluation of $\langle X_k \rangle(t)$, using the information obtained in step 2. $\langle X_k \rangle(t)$ will depend on the matrix elements $\langle A_i| X_k|A_j \rangle$, which, in the case of a free particle, will be at most linear functions of time. For the purposes of this investigation, we specialize to the case in which the $\langle A_i| X_k|A_j \rangle$ are simply constants, because we are interested in the motion induced by the non-linearity, and not the linear motion of the system as a whole.

In ref. \cite{ChodosCooper2025} we carried out this sequence of steps analytically for a very simple case, in which $\kappa$ is a constant and $\tau=0$. The result was an elliptic (or possibly circular) trajectory in which the particle moves harmonically about the center. A full analytic treatment of other cases has so far not proved possible. In this paper we perturb about the simple solution and also perform a numerical investigation of the properties of more general trajectories.

The $\tau-\delta$ equations follow directly from the equations of motion. They are
\begin{equation}
    \frac{d\kappa}{dt} = - 2 (b+\mu) \tau
    \label{eq:EoM_kappa_t}
\end{equation}
and 
\begin{equation}
    \frac{d\tau}{dt} = \mu (N^2-\tau^2) + 4 be^\kappa
    \label{eq:EoM_tau_t}
\end{equation}
where $\delta=e^\kappa$. These equations admit a first integral:
\begin{equation}
    N^2 = \tau^2 + 4 e^\kappa + c e^{p\kappa},
    \label{eq:N_integral}
\end{equation}
where $p=\frac{\mu}{b+\mu}$ and $c$ is a constant of integration. As shown in \cite{ChodosCooper2025}, the Schwarz inequality requires $c$ to be non-negative. $\kappa$ satisfies the equation of motion of a classical particle in one dimension, with momentum $\tau$, moving in the potential $V(\kappa) = 4 e^\kappa + c e^{p \kappa}$. As in previous work, we choose $p<0$, so that the motion of $\kappa$ is bounded and periodic. 

It is convenient to write our equations in $2\times 2$ matrix form. We define time-dependent functions $\chi_i (t)$ and $\xi_i(t), i =1,2,$ via
\begin{equation}
    | \psi_1 \rangle = e^{p \kappa/4} e^{-\frac{i}{2}N\lambda t} \left[ \chi_1 |A_1 \rangle + \chi_2 |A_2\rangle \right]
    \label{eq:psi_1_in_chi_xi}
\end{equation}
and 
\begin{equation}
    | \psi_2 \rangle = e^{p \kappa/4} e^{\frac{i}{2}N\lambda t} \left[ \xi_1 |A_1 \rangle + \xi_2 |A_2\rangle \right]
        \label{eq:psi_2_in_chi_xi}
\end{equation}
As noted above, the $|A_i\rangle$ are 2 linearly independent vectors that evolve according to the underlying Hamiltonian $H$, and are normalized as $\langle A_i|A_j\rangle = \delta_{ij}$.

From the equations of motion, the matrix 
\begin{equation}
    Z = \begin{pmatrix}
        \chi_1 & \chi_2 \\ \xi_1 & \xi_2
    \end{pmatrix}
    \label{eq:Z_matrix_def}
\end{equation}
then satisfies
\begin{equation}
    \dot Z = M Z
    \label{eq:EoM_Z}
\end{equation}
with
\begin{equation}
    M = \begin{pmatrix}
        \frac{N}{2}(\mu+i\lambda) & -i(g+\lambda)\gamma' \\ -i(g^*-\lambda) \gamma'^* & -\frac{N}{2}(\mu + i \lambda)
    \end{pmatrix}.
    \label{eq:M_matrix_def}
\end{equation}
Here $\gamma = \langle \psi_2 |\psi_1 \rangle = \gamma' e^{-i N \lambda t},$ and
\begin{equation}
    \gamma' = e^{\frac{\kappa}{2}\left[ 1- \frac{ia}{b+\mu} \right]}= e^{\frac{\kappa}{2}\zeta}.
    \label{eq:gamma_and_zeta}
\end{equation}
We note that $M$ is not anti-Hermitian, which might have been expected as a necessary requirement to ensure unitary evolution. However, the key property that guarantees the absence of any exponentially growing solutions is principle 1. above, which is obeyed by our system of equations. In appendix \ref{sec:app_B}, we argue that what is or is not Hermitian can depend on one’s point of view.

One can demonstrate that the boundary conditions
\begin{equation}
    \langle \psi_1|\psi_1 \rangle + \langle \psi_2|\psi_2 \rangle = N, \quad \langle \psi_1|\psi_1 \rangle - \langle \psi_2|\psi_2 \rangle = \tau,  \quad\text{ and }\quad \langle \psi_2|\psi_1 \rangle = \gamma.
    \label{eq:boundary_conditions}
\end{equation}
are encoded in the matrix $ZZ^\dagger$, which must take the form
\begin{equation}
    ZZ^\dagger = \frac{1}{2}e^{-\frac{p\kappa}{2}} \begin{pmatrix}
        (N+\tau) & 2\gamma' \\ 2 \gamma'^* & (N-\tau)
    \end{pmatrix}.
    \label{eq:boundary_applied_to_ZZ}
\end{equation}
Following from its definition for the expectation value of an operator $X$ is
\begin{equation}
    \langle X \rangle = \frac{1}{N} e^{\frac{p \kappa}{2}} \sum_{i,j} \left[ \langle A_i |X|A_j \rangle \left( Z^\dagger Z \right)_{ij} \right].
    \label{eq:X_matrix_elements}
\end{equation}
But
\begin{equation}
    \operatorname{Tr} Z^\dagger Z = \operatorname{Tr} ZZ^\dagger = N e^{-\frac{p\kappa}{2}}
\end{equation}
and 
\begin{equation}
    \operatorname{det} Z^\dagger Z = \operatorname{det} ZZ^\dagger = \frac{e^{-p\kappa}}{4} \left[ N^2 - \tau^2 - 4e^\kappa \right] = \frac{c}{4}.
\end{equation}
We can therefore represent $Z^\dagger Z$ as
\begin{equation}
    Z^\dagger Z = \frac{N}{2} e^{-\frac{p\kappa}{2}}\left[ 1+\vec u \cdot \vec \sigma \right].
    \label{eq:ZZ_dagger_in_u}
\end{equation}
Using $\operatorname{det} \left[ 1+\vec u \cdot \vec \sigma \right]=1-\vec u ^2 $, we have
\begin{equation}
    \vec u^2 = 1 - \frac{ce^{p\kappa}}{N^2}.
    \label{eq:vec_u_squared}
\end{equation}
Applying this to the case of a position operator of a free particle, and using our assumption that the matrix elements $\langle A_i | X_k |A_j \rangle \equiv (X_{ji})_k$ are constants (note the flipped indices), we similarly expand $(X_{ji})_k=X_{0k} \delta_{ij} + \vec X_k \cdot \vec \sigma_{ji}$, which leads to
\begin{equation}
    \langle X_k \rangle = \frac{1}{2} \operatorname{Tr} \left[ X_{0k} + \vec X_k \cdot \vec \sigma \right] \left[ 1+ \vec u \cdot \vec \sigma \right] = X_{0k} + \vec X_k \cdot \vec u. 
\end{equation}
The $X_{0k}$ are just constants, reflecting the freedom to choose the origin of $X_k$, so if we let
\begin{equation}
    \langle X'_k\rangle = \langle X_k \rangle - X_{0k},
    \label{eq:x_centered}
\end{equation}
we simply have 
\begin{equation}
    \langle X'_k \rangle = \vec X_k \cdot \vec u.
    \label{eq:trajectory_in_u}
\end{equation}
Here the $\vec X_k$ are constant; all the time dependence is in $\vec u$. Because we know $\vec u^2$, there are only 2 undetermined functions of time to determine the trajectory. 

\section{Some Analytic Results}
\label{sec:analytic_res}
\subsection{$c=0$ is ordinary quantum mechanics}
When $c=0$, it follows from the equation \eqref{eq:N_integral} and the definitions \eqref{eq:tau_and_gamma_def} that we also have $\mathcal{S}=0$, where
\begin{equation}
    \mathcal{S} = \langle \psi_1 | \psi_1\rangle \langle \psi_2 | \psi_2\rangle - |\langle \psi_1 | \psi_2\rangle|^2.
    \label{eq:Schwartz}
\end{equation}
This implies that $|\psi_2 (t) \rangle = A(t) |\psi_1 (t)\rangle$ for some $A(t)$. (We will determine $A(t)$ below). Since there is effectively only one state vector, we suspect that the theory reduces to ordinary quantum mechanics. We now demonstrate this explicitly.
With $c=0$, we have $N^2 = \tau^2 + 4 \delta = \tau^2 +4 e^\kappa$. The $\tau-\delta$ equations are
\begin{equation}
    \frac{d\kappa}{dt} = -2 (b+\mu)\tau
\end{equation}
and
\begin{equation}
    \frac{d \tau}{dt} = \mu (N^2 - \tau^2) + 4 be^\kappa,
\end{equation}
whose solution in this case is
\begin{equation}
    \tau = -N \tanh Ns
\end{equation}
and
\begin{equation}
    4 e^\kappa = N^2 - \tau^2 = \frac{N^2}{\cosh^2 Ns},
\end{equation}
where $s = -(b+\mu) t$.
We can determine $A(t)$ from the equation $\langle \psi_1 | \psi_2 \rangle = \gamma^*$ to be
\begin{equation}
    A(t) = \frac{2\gamma^*}{N+\tau}.
\end{equation}
 It then follows that
 \begin{equation}
     1+|A|^2 = 1 + \frac{4e^\kappa}{(N+\tau)^2} = 1 +\frac{N^2-\tau^2}{(N+\tau)^2} = \frac{2N}{(N+\tau)}=2 \cosh Ns~e^{Ns}.
 \end{equation}
 Now we substitute $|\psi_2 (t) \rangle = A(t) |\psi_1 (t)\rangle $ into the equation of motion for $|\psi_1(t)\rangle$. We obtain a first-order equation for $|\psi_1 (t)\rangle$:
 \begin{equation}
     i \frac{d |\psi_1 (t)\rangle}{dt} = \left[ H + \frac{1}{2}(N-\tau)(i \mu + g + \lambda) \right] |\psi_1(t)\rangle,
 \end{equation}
 where we have used
 \begin{equation*}
     \frac{2|\gamma|^2}{N+\tau} = \frac{1}{2}(N-\tau).
 \end{equation*}
 We can write the bracket on the right-hand side as $\frac{d\xi}{dt}$, where
 \begin{equation}
     \xi = Ht + \frac{1}{2}N (i\mu +g +\lambda)t +\frac{1}{4}\frac{1}{(b+\mu)}(i\mu + g + \lambda) \kappa \equiv Ht+\eta.
 \end{equation}
Hence
\begin{equation}
    |\psi_1(t) \rangle = e^{-i\xi} |\psi_0\rangle,
\end{equation}
where $|\psi_0 \rangle$ is a constant vector.

Now let $X$ be an arbitrary operator, and compute
\begin{equation}
    \langle X\rangle = \frac{1}{N} \left[ \langle \psi_1 (t)|X|\psi_1 (t) \rangle + \langle \psi_2 (t) | X|\psi_2 (t)\rangle \right] = \frac{1}{N} \left[ 1+|A|^2 \right] \langle \psi_1 (t)|X| \psi_1 (t)\rangle.
\end{equation}
But
\begin{equation}
    \langle \psi_1 (t) |X|\psi_1 (t)\rangle = \langle \psi_0 | e^{iHt} X e^{-i Ht}|\psi_0\rangle e^{i(\eta^* -\eta)}
\end{equation}
and we find
\begin{equation}
    i(\eta^*-\eta) = (b+\mu)Nt + \frac{\kappa}{2},
\end{equation}
so that
\begin{equation}
    e^{i(\eta^* - \eta)} = \left[ e^{-Ns} \right] \left( \frac{N}{2}\frac{1}{\cosh Ns} \right).
\end{equation}
Multiplying everything together, we have
\begin{equation}
    \langle X(t)\rangle = \langle \psi_0 |e^{i H t} X e^{-iHt}|\psi_0\rangle,
\end{equation}
which is just what we expect from ordinary quantum mechanics. (The requirement that $\langle \psi_1 | \psi_1 \rangle + \langle \psi_2 | \psi_2 \rangle = N$ fixes $\langle \psi_0 | \psi_0 \rangle = 1.$) 

\subsection{The case $b+\mu = 0$}
Here we present a solution for $|\psi_1 \rangle$ and $|\psi_2 \rangle$ in the case $b+\mu = 0$. As remarked in reference \cite{ChodosCooper2025}, this case is not covered by our previous analysis, because we cannot define $s=-(b+\mu)t$, which is the evolution parameter for the $(\kappa, \tau)$ system, and so the relation
\begin{equation*}
    N^2 = \tau^2 + 4e^\kappa + c e^{p\kappa}
\end{equation*}
does not hold.

As noted in the reference \cite{ChodosCooper2025}, when $b+\mu = 0$ the solutions for $\kappa$ and $\tau$ are
\begin{equation}
    \delta = e^\kappa = \frac{N^2}{4}- \omega_0^2
\end{equation}
and
\begin{equation}
    \tau = -2\omega_0 \tanh{2\omega_0 bt}.
    \label{eq:tau_b_and_mu_0}
\end{equation}
Here $\omega_0$ is a constant of integration limited by the condition $\omega_0^2 \leq \frac{N^2}{4}$.

We plug these values into the equations for $\chi$ and $\xi$. To keep the analysis simple, we also set the parameters $a$ and $\lambda$ to zero, so what follows is only a particular solution to our system of equations in the case $b+\mu=0$. We have
\begin{equation}
    \begin{aligned}
        \dot\chi &= -\frac{b}{2}(N-\tau) \chi + b e^{\kappa/2} \xi \\
        \dot\xi &= \frac{b}{2}(N+\tau) \xi - b e^{\kappa/2} \chi
    \end{aligned}
\end{equation}
and note that if
\begin{equation}
    f = -\frac{1}{2}\ln{\cosh{2\omega_0 bt}},
\end{equation}
then
\begin{equation}
    \dot f = \frac{b}{2}\tau.
\end{equation}
We make the substitutions $\chi = e^f \chi'$ and $\xi = e^f \xi'$ and find
\begin{equation}
    \dot \chi' = -\frac{b}{2} N\chi' + b e^{\kappa/2} \xi',
\end{equation}
and
\begin{equation}
    \dot \xi' = \frac{b}{2} N\xi' - b e^{\kappa/2} \chi'.
\end{equation}
All the coefficients in these equations are constants, so the equations are easily solved. We have two independent solution. Ignoring overall normalization, these are:
\begin{equation}
    \chi_+' = \cos{\theta_+} ~e^{b \omega_0 t}, \quad \xi'_+ = \sin{\theta_+} ~e^{b\omega_0 t} 
\end{equation}
with $\tan{\theta_+} = \frac{e^{\kappa/2}}{\frac{N}{2}-\omega_0} = \sqrt{\frac{\frac{N}{2}+\omega_0}{\frac{N}{2}-\omega_0}}$, and
\begin{equation}
    \chi_-' = \cos{\theta_-} ~e^{-b \omega_0 t}, \quad \xi'_- = \sin{\theta_-} ~e^{-b\omega_0 t} 
\end{equation}
with $\tan{\theta_-} = \frac{e^{\kappa/2}}{\frac{N}{2}+\omega_0} = \sqrt{\frac{\frac{N}{2}-\omega_0}{\frac{N}{2}+\omega_0}}$. The normalization is determined by imposing the conditions
\begin{equation}
    \langle \psi_1|\psi_1 \rangle + \langle \psi_2|\psi_2 \rangle = N, \quad \langle \psi_1|\psi_1 \rangle - \langle \psi_2|\psi_2 \rangle = \tau,  \quad\text{ and }\quad \langle \psi_2|\psi_1 \rangle = e^{\frac{\kappa}{2}},
\end{equation}
(we are keeping everything real and hence ignoring possible phases). Let
\begin{equation}
    \mathcal{N}(t) = \sqrt\frac{N}{2} \left( \cosh{2\omega_0 bt} \right)^{-\frac{1}{2}}.
\end{equation}
Then we have
\begin{equation}
    |\psi_1 \rangle = \mathcal{N}(t) \left[ \cos\theta_+ ~e^{b\omega_0t}|A_1 \rangle+ \cos{\theta_-} ~e^{-b\omega_0 t} |A_2\rangle \right]
\end{equation}
and
\begin{equation}
    |\psi_2 \rangle = \mathcal{N}(t) \left[ \sin\theta_+ ~e^{b\omega_0t}|A_1 \rangle+ \sin{\theta_-} ~e^{-b\omega_0 t} |A_2\rangle \right],
\end{equation}
where $\langle A_i | A_j \rangle = \delta_{ij},$ and
\begin{equation}
    \cos{2\theta_+} = - \cos{2\theta_-}=\frac{-2\omega_0}{N}
\end{equation}
and
\begin{equation}
    \sin{2\theta_+} = \sin{2\theta_-}=\frac{2}{N}e^{\kappa/2}.
\end{equation}
It is noteworthy (and necessary) that these conditions on $\theta_+$ and $\theta_-$ are compatible with the earlier conditions
\begin{equation}
    \tan \theta_\pm = \frac{e^{\kappa/2}}{\frac{N}{2}\mp \omega_0}.
\end{equation}
This can be checked using the identity
\begin{equation}
    \tan \theta = \frac{\sin 2\theta}{1+\cos 2\theta}.
\end{equation}
Just as with the "simple" solution in reference \cite{ChodosCooper2025}, there appear to be more conditions than there are parameters to satisfy them, but (miraculously?) a solution exists.

We now compute the trajectory associated with this solution. For any operator $X$ the expectation value is
\begin{equation}
    \langle X\rangle = \varphi \left[ e^{2b\omega_0 t} \langle A_1|X|A_1\rangle + e^{-2b\omega_0 t} \langle A_2|X|A_2 \rangle + 2 \cos{(\theta_+ - \theta_-) \operatorname{Re}}\,\langle A_1|X|A_2 \rangle \right],
\end{equation}
where $\varphi = \frac{1}{2\cosh{2\omega_0 bt}}$. Despite the appearance of exponential behavior in the square bracket, the expectation value remains bounded due to the time dependence in $\varphi$.

We take $\vec X$ to be a position operator and assume its matrix elements are constants. Then the expression for $\langle \vec X \rangle$ can be rearranged as
\begin{equation}
    \langle \vec X\rangle - \vec X_0 \equiv \vec X' (t) = \vec V \tanh{2\omega_0 b t} + \vec W \frac{1}{\cosh{2\omega_0 bt}}
\end{equation}
for suitably defined constants $\vec X_0, \vec V$ and $\vec W$.
Using
\begin{equation}
    \tanh^2{2\omega_0 bt} + \frac{1}{\cosh^2{2\omega_0bt}}=1,
\end{equation}
we have
\begin{equation}
    \left( \vec V \times \vec X' \right)^2 + \left( \vec W \times \vec X' \right)^2 = \left( \vec V \times \vec W \right)^2.
\end{equation}
Using
\begin{equation}
    \left( \vec A \times \vec B \right)^2 = A^2 B^2 - \left( \vec A \cdot \vec B \right)^2,
\end{equation}
and choosing the plane spanned by $\vec V$ and $\vec W$ to be the $x-y$ plane, we can write this as
\begin{equation}
    \begin{pmatrix}
        x' & y'
    \end{pmatrix} \begin{pmatrix}
        v_2^2+w^2_2 & - (v_1 v_2 + w_1w_2) \\ -(v_1v_2+w_1w_2) & v_1^2 + w_1^2
    \end{pmatrix} \begin{pmatrix}
        x' \\ y'
    \end{pmatrix} = \left( \vec V \times \vec W \right)^2.
\end{equation}
The matrix on the lhs is real and symmetric and can therefore be diagonalized by an orthogonal transformation. Its trace equals $V^2 + W^2$ and is positive. Its determinant is $\left( \vec V \times \vec W \right)^2$ and is also positive. Therefore, both eigenvalues are positive, so the equation describes an ellipse in the plane spanned by $\vec V$ and $\vec W$. As $t$ goes from $-\infty$ to $\infty$, $\vec X'$ goes from $-\vec V$ to $\vec V$ along the curve. The particle motion is bounded, but, unlike in examples when $(b+\mu) \neq 0$, it is not quasi-periodic.

\subsection{Analytic expression for small oscillation amplitude}
\label{sec:small_osc_analytic}
We employ the formalism outlined in section \ref{sec:model_review}. We want to obtain an expression for Z, as defined in equation \eqref{eq:Z_matrix_def}, valid to first order in small oscillations about the solution $e^{\kappa_0}=\frac{-\mu}{4b}N_0^2$ and $\tau=0$. We write
\begin{equation}
    \kappa = \kappa_0+R\eta,
\end{equation}
where $R$ will be taken small. We find $N^2 = N_0^2 + \Delta$, where $\Delta = \tau^2 + 2 (1-p) e^{\kappa_0} R^2 \eta^2$, and the equations of motion
\begin{equation}
    \frac{d^2 \eta}{ds^2} = -\omega'^2\eta,
\end{equation}
where
\begin{equation}
    \omega'^2 = -2pN_0^2.
    \label{eq:omega'}
\end{equation}
Therefore, $\eta$ has the general solution (up to an overall constant, which we absorb in $R$)
\begin{equation}
    \eta = \cos{(\omega t - \theta_0)}
\end{equation}
and
\begin{equation}
    \tau = \frac{-1}{2(b+\mu)}\frac{d\kappa}{dt}=\frac{\omega R}{2(b+\mu)}\sin{(\omega t - \theta_0)},
    \label{eq:tau_approx}
\end{equation}
where $\omega = (b+\mu) \omega'$ and $\theta_0$ is a constant. Note that $\Delta = \frac{\omega'^2}{4}R^2$.
We expand $M$ and $Z$ to the first order in $R$:
\begin{equation}
    M = M_0 + R M_1, \qquad Z = Z_0 + RZ_1,
\end{equation}
which leads to the equations of motion
\begin{equation}
    \dot Z_0 = M_0 Z_0, \qquad \dot Z_1 = M_0Z_1 + M_1Z_0.
\end{equation}
The solutions are
\begin{equation}
    Z_0 = e^{M_0 t} Z_0(0) \qquad Z_1 = e^{M_0 t} \left[ W_1 + F(t) \right] Z_0(0).
\end{equation}
Here
\begin{equation}
    F(t) = \int_0^t e^{-M_0t'} M_1 (t') e^{M_0 t'} dt'
\end{equation}
and $W_1$ is a constant (matrix) of integration.

Since both $M_0$ and $M_1$ are traceless, we can expand them as
\begin{equation}
    M_0 = \vec a \cdot \vec \sigma, \qquad M_1 = e^{\kappa_0/2} \left( \frac{-i\eta}{2} \right) \vec b \cdot \vec \sigma.
\end{equation}
We will give explicit expressions for $\vec a$ and $\vec b$ below.

We use the identities
\begin{equation}
    e^{\vec a \cdot \vec\sigma t} = \cos{\omega_1 t} + \frac{\vec a \cdot \vec \sigma}{\omega_1} \sin{\omega_1 t},
\end{equation}
where $\omega_1^2 = -\vec a^2$, and
\begin{equation}
    e^{-\vec a \cdot \vec\sigma t}\left(\vec b \cdot \vec \sigma \right) e^{\vec a \cdot \vec\sigma t} = \left[ \vec b \cdot \vec \sigma + \frac{\vec a \cdot \vec b}{\omega_1^2} \vec a \cdot \vec \sigma\right] \cos{2\omega_1 t} + \frac{i}{\omega_1} \left( \vec b \times \vec a\cdot \vec \sigma \right) \sin{2\omega_1 t} - \frac{\vec a \cdot \vec b}{\omega_1^2} \vec a \cdot \vec \sigma.
\end{equation}
The last equation can be abbreviated as 
\begin{equation}
    e^{-\vec a \cdot \vec\sigma t}\left(\vec b \cdot \vec \sigma \right) e^{\vec a \cdot \vec\sigma t} = \left( \vec k_1\cos{2\omega_1 t} + \vec k_2 \sin{2\omega_1 t} + \vec k_3 \right) \cdot \vec \sigma,
\end{equation}
with the obvious definitions of the $\vec k_i$ in terms of $\vec a$ and $\vec b$.

Inserting this in the expression for $F(t)$, and remembering the extra factor of $\eta$ in the definition of $M_1$, we see that the functions we have to integrate are $\cos{2\omega_1t}\cos{(\omega t - \theta_0)},\,\sin{2\omega_1t}\cos{(\omega t - \theta_0)}$, and $\cos{(\omega t-\theta_0)}$. Using
\begin{equation}
    \cos{A}\cos{B} = \frac{1}{2} \left[ \cos{(A+B)}+ \cos{(A-B)} \right]
\end{equation}
and
\begin{equation}
    \sin{A}\cos{B} = \frac{1}{2} \left[ \sin{(A+B)}+ \sin{(A-B)} \right],
\end{equation}
we obtain
\begin{equation}
    F(t) = \frac{-i e^{\kappa_0/2}}{2} \left[J_1 \vec k_1 + J_2 \vec k_2 + J_3 \vec k_3 \right] \cdot \vec \sigma,
\end{equation}
where
\begin{equation}
    \begin{aligned}
        J_1 &= \frac{1}{2} \left[ \frac{\sin \left[ (\omega + 2 \omega_1)t-\theta_0 \right]}{(\omega + 2 \omega_1)} + \frac{\sin \left[ (\omega - 2 \omega_1)t-\theta_0 \right]}{(\omega - 2 \omega_1)} + \left( \frac{2\omega}{\omega^2 - 4 \omega_1^2} \right) \sin \theta_0 \right], \\
        J_2 &= -\frac{1}{2} \left[ \frac{\cos \left[ (\omega + 2 \omega_1)t-\theta_0 \right]}{(\omega + 2 \omega_1)} - \frac{\cos \left[ (\omega - 2 \omega_1)t-\theta_0 \right]}{(\omega - 2 \omega_1)} + \left( \frac{4\omega_1}{\omega^2 - 4 \omega_1^2} \right) \cos \theta_0 \right], \\
        J_3 &= \frac{1}{\omega} \left[ \sin{(\omega t-\theta_0)} + \sin \theta_0 \right].
    \end{aligned}
\end{equation}
The values of $\vec a$ and $\vec b$ in terms of the original parameters are
\begin{equation}
    \begin{aligned}
        a_1 &= -i e^{\kappa_0/2} \left[ a\,\cos{\tilde{\theta}_0} + (b-i\lambda) \sin{\tilde{\theta}_0} \right], \\
        a_2 &= -i e^{\kappa_0/2} \left[ a\,\sin{\tilde{\theta}_0} - (b-i\lambda) \cos{\tilde{\theta}_0} \right], \\
        a_3 &= \frac{N_0}{2}(\mu + i\lambda),
    \end{aligned}
\end{equation}
where
\begin{equation}
    \tilde{\theta}_0 = \frac{a \kappa_0}{2(b+\mu)}.
\end{equation}

We calculate
\begin{equation}
    \omega_1^2 = - \vec a^2 = e^{\kappa_0} \left[ a^2 + b (b+\mu) - \frac{\lambda^2}{\mu}(b+\mu) \right].
    \label{eq:omega_1}
\end{equation}
With our choice of parameters, all 3 terms in this expression are positive, so $\omega_1$ is real. We also have 
\begin{equation}
\begin{aligned}
    b_1 &= c_1 \cos{\tilde{\theta}_0} - c_2 \sin{\tilde{\theta}_0}, \\
    b_2 &= c_2 \cos{\tilde{\theta}_0} + c_1 \sin{\tilde{\theta}_0}, \\
    b_3 &= 0.
    \end{aligned}
\end{equation}
Here
\begin{equation}
    c_1 = a \left[ 1- \frac{i(\lambda + i b)}{b+\mu} \right], \qquad c_2 = \frac{a^2}{b+\mu}+ i (\lambda+i b).
\end{equation}
This completes the specification of all the ingredients that go into $Z(t)$ to the first order in $R$, except for the boundary conditions, which we now discuss.

We make use of the fact that
\begin{equation}
    Z Z^\dagger = \frac{e^{-\frac{p\kappa}{2}}}{2} \begin{pmatrix}
        (N+\tau) & 2\gamma' \\ 2 \gamma'^* & (N-\tau)
    \end{pmatrix}.
\end{equation}
Following the previous work, at zeroth order in $R$ we choose $\chi_2=0$ at $t=0$. Suitable choice for the other parameters in $Z_0(0)$ are
\begin{equation}
    \begin{aligned}
        \chi_1 &= \sqrt\frac{N_0}{2} e^{-\frac{p\kappa_0}{4}}, \\
        \xi_1 &= e^{-\frac{p\kappa_0}{4}} e^{\frac{\zeta^* \kappa_0}{2}} \sqrt \frac{2}{N_0}, \\
        \xi_2 &= e^{\frac{p\kappa_0}{4}}\sqrt \frac{c}{2N_0}.
    \end{aligned}
\end{equation}
It remains to determine the matrix $W_1$. At $t=0$, to the first order in $R$, we have
\begin{equation}
    Z = \left( 1+R W_1 \right) Z_0(0).
\end{equation}
So, we must compare $\left(W_1 Z_0(0)Z_0(0)^\dagger + Z_0(0) Z_0(0)^\dagger W_1^\dagger \right)$ to the expansion of $ZZ^\dagger$ to the first order. For simplicity we shall now choose $\theta_0 = \pi/2$, so that
\begin{equation}
    \eta(0) = 0 \quad \text{ and } \quad \tau(0) = -\frac{R \omega}{2(b+\mu)}.
\end{equation}
This then gives
\begin{equation}
    W_1 Z_0(0)Z_0(0)^\dagger + Z_0(0) Z_0(0)^\dagger W_1^\dagger = - e^{-\frac{p\kappa_0}{2}} \frac{\omega}{4(b+\mu)}\sigma_3.
\end{equation}
We observe that
\begin{equation}
    Z_0(0)Z_0(0)^\dagger = \frac{e^{\frac{p\kappa_0}{2}}}{2} N_0 \left[ 1+Q \right],
\end{equation}
where
\begin{equation}
    Q = \begin{pmatrix}
        0 & \frac{2}{N_0}e^{\frac{\kappa_0}{2}\zeta} \\ \frac{2}{N_0} e^{\frac{\kappa_0}{2}\zeta^*} & 0
    \end{pmatrix},
\end{equation}
satisfying
\begin{equation}
    Q^2 = \frac{4}{N_0^2} e^{\kappa_0},
\end{equation}
so that 
\begin{equation}
    \left[ Z_0(0) Z_0(0)^\dagger \right]^{-1} = 2 N_0 e^{-\frac{p\kappa_0}{2}} \frac{1}{c} (1-Q).
\end{equation}
We define
\begin{equation}
    W_1 = V_1 \left[ Z_0(0) Z_0(0)^\dagger \right]^{-1},
\end{equation}
so that
\begin{equation}
    V_1 + V_1^\dagger = - e^{\frac{-p\kappa_0}{2}} \frac{\omega}{4(b+\mu)} \sigma_3.
\end{equation}
And it, therefore, suffices to take
\begin{equation}
    V_1 = -e^{-\frac{p\kappa_0}{2}}\frac{\omega}{8(b+\mu)} \sigma_3.
\end{equation}
Putting everything together then gives all the ingredients necessary for an analytic solution to the small oscillation problem to the first order.

%%%%%%%%%%%%%%%%%%%%%%%%%%%%%%%%%%%%%%%%%%
\section{Floquet structure of the periodic sector}
\label{sec:floquet_structure}

For $p<0$, $c>0$, and $N^2>V_{\min}$, the nonlinear driver $(\kappa,\tau)$ is periodic, with period $T_\kappa$.  As a consequence, the coefficient matrix in Eq.~\eqref{eq:EoM_Z} obeys
\begin{equation}
    \dot Z=M[\kappa(t)]Z,
    \qquad M(t+T_\kappa)=M(t).
\end{equation}
Let $\mathcal Z(t)$ be the fundamental matrix solution normalized by $\mathcal Z(0)=\mathbb{I}$.  The monodromy matrix
\begin{equation}
    \mathcal M=\mathcal Z(T_\kappa)
\end{equation}
then determines the evolution from one driver period to the next \cite{Floquet1883,Chicone2024}.

In the present model physical periodic solutions are elliptic, which follows directly from the boundary constraints (\ref{eq:boundary_conditions}). For $c>0$, the exact identity established earlier, $\det(Z^\dagger Z)=c/4$, guarantees that the initial state matrix $Z_0:=Z(0)$ is invertible. At some arbitrary time $t$ this state matrix can be expressed through the fundamental matrix solution $\mathcal Z (t)$ as $Z(t)=\mathcal Z(t)Z_0$. Evaluating it after one driver period and using $(ZZ^\dagger)(T_\kappa)=(ZZ^\dagger)(0)$, we obtain
\begin{equation}
    \mathcal M G_0\mathcal M^\dagger=G_0,
    \qquad
    G_0=Z_0Z_0^\dagger>0.
    \label{eq:periodic_gram_floquet}
\end{equation}
Thus, although the coefficient matrix $M$ is not anti-Hermitian, its one-period monodromy $\mathcal M$ preserves the positive-definite Hermitian form determined by metric $\eta_G=G_0^{-1}$. The monodromy is, therefore, pseudo-unitary with respect to this metric, in the terminology of Ref.~\cite{Mostafazadeh2004}. Unlike pseudo-unitarity with an indefinite metric, which may admit non-unimodular eigenvalues, the positive-definite case implies similarity to a unitary matrix. This allows us to introduce a unitary evolution operator
\begin{equation}
    U_F:=Z_0^{-1}\mathcal M Z_0.
\end{equation}
Indeed, by applying the equation~\eqref{eq:periodic_gram_floquet}, we have
\begin{equation}
    U_F U_F^\dagger = Z_0^{-1} \mathcal{M} Z_0 \left( Z_0^{-1} \mathcal{M} Z_0 \right)^\dagger = Z_0^{-1} \underbrace{\mathcal{M} Z_0 Z_0^\dagger \mathcal{M}^\dagger}_{Z_0Z_0^\dagger }(Z_0^{-1})^\dagger=\mathbb{I}.
\end{equation}
Additionally, since $\operatorname{Tr}M(t)=0$, the fundamental matrix has its determinant equal to unity, and hence $\det U_F=\det\mathcal M=1$. In this sense, the one-period evolution is similar to an $SU(2)$ transformation, $U_F\in SU(2)$. By Floquet periodicity the solution from one driver period to the next can be obtained by $Z(t+T_\kappa)=Z(t)U_F$. Hence, for a given periodic solution, each complete cycle induces the same unitary transformation on the two-component coefficient spinors, $\xi_i$ and $\chi_i$, even though the evolution of the full state vectors in the original Hilbert-space representation is non-unitary.

This stroboscopic evolution has a natural geometric interpretation in the Bloch representation. In the (in general) freely evolving orthonormal basis $\{|A_1(t)\rangle,|A_2(t)\rangle\}$, the physical density operator defined in Eq.~\eqref{eq:density_matrix_operator} is represented by the two-level density matrix
\begin{equation}
    \rho(t)
    =\frac{e^{p\kappa/2}}{N}(Z^\dagger Z)^T
    =\frac12\left(
        \mathbb{I}+\mathbf B(t)\cdot\boldsymbol\sigma
    \right),
    \qquad
    \mathbf B=J\vec u,
    \quad J=\operatorname{diag}(1,-1,1).
    \label{eq:bloch_prior_to_recurrence}
\end{equation}
The Bloch vector $\mathbf B$ is thus equivalent to the previously introduced $\vec u$, expressed in the standard coordinates of the Bloch ball \cite{BengtssonZyczkowski2017}. Its radial evolution is governed by the exact nonlinear constraint
\begin{equation}
    |\mathbf B(t)|^2
    =1-\frac{ce^{p\kappa(t)}}{N^2},
\end{equation}
as follows from Eq.~\eqref{eq:vec_u_squared}.

Since the coefficient matrix $M(t)$ is $T_\kappa$-periodic, the right action of $U_F$ induces the exact stroboscopic density-matrix evolution\footnote{Defining $W_F=U_F^T$, the stroboscopic density-matrix evolution takes the standard unitary form $\rho(t+T_\kappa)=W_F\rho(t)W_F^\dagger$. The transpose arises solely from our convention for the coefficient matrix $Z$, in which the two state vectors are represented by rows, so that the reduced density matrix is proportional to $(Z^\dagger Z)^T$ rather than $Z^\dagger Z$.}
\begin{equation}
    \rho(t+T_\kappa)
    =U_F^T\rho(t)U_F^*,
    \qquad
    \mathbf B(t+T_\kappa)
    =Q_F^B\mathbf B(t),
    \quad Q_F^B\in SO(3).
    \label{eq:bloch_rotation}
\end{equation}
Here $Q_F^B$ is the rotation induced by the conjugation action of $U_F^T\in SU(2)$ on the two-level density matrix \cite{BengtssonZyczkowski2017}. Thus, while the nonlinear driver controls the periodic radial motion, the Floquet map determines the reorientation of the Bloch trajectory from one driver period to the next.

The angle of this stroboscopic rotation is fixed by the eigenphases of the monodromy. Since $\mathcal M$ is similar to an $SU(2)$ matrix, its eigenvalues can be written as $e^{\pm i\theta_{F}}$, where the principal Floquet phase and ratio are defined by
\begin{equation}
    \theta_{F}
    =\arccos\left(
        \frac{\operatorname{Tr}\mathcal M}{2}
    \right)\in[0,\pi],
    \qquad
    r_{\mathrm{pr}}
    =\frac{\theta_{F}}{2\pi}
    \in[0,1/2].
    \label{eq:floquet_minimal_ratio}
\end{equation}
The corresponding Bloch rotation angle is $2\theta_{F}$ modulo $2\pi$. These quantities characterize the period-to-period reorientation and provide the basis for the recurrence conditions discussed below.

The principal Floquet phase $\theta_F$ uniquely specifies the unordered pair of monodromy eigenvalues, but does not distinguish their continuously evolving branches when the model parameters are varied. Indeed, the eigenphases are defined modulo $2\pi$, and the two Floquet multipliers may be interchanged. To follow the evolution across successive Floquet zones, we therefore introduce a lifted eigenphase. Consider, for example, a parameter-dependent family $\mathcal M(a)$. On an interval where its eigenvalues remain distinct, we choose one multiplier continuously and write
\begin{equation}
e^{i\Theta_F(a)}
    \in\operatorname{spec}\mathcal M(a),
    \qquad \Theta_F(a)\in\mathbb R.
    \label{eq:lifted_phase_definition}
\end{equation}
The lift is fixed by specifying the selected multiplier and its phase at a reference point $a_0$. The corresponding lifted Floquet ratio is
\begin{equation}
    r(a):=\frac{\Theta_F(a)}{2\pi}.
    \label{eq:lifted_floquet_ratio}
\end{equation}
It is related to the principal ratio by
\begin{equation}
    r_{\mathrm{pr}}(a)
    =\operatorname{dist}\bigl(r(a),\mathbb Z\bigr).
    \label{eq:lifted_to_principal}
\end{equation}
Thus, $r(a)$ follows a chosen continuous branch, whereas $r_{\mathrm{pr}}(a)$ is folded into the fundamental interval $[0,1/2]$. Different choices of the eigenvalue branch or its initial phase correspond to transformations like $r\mapsto-r$ or $r\mapsto r+k$, $k\in\mathbb Z$, without changing the monodromy spectrum. This is analogous to the quasienergy-zone convention of standard periodically driven quantum systems \cite{Shirley1965,Sambe1973}, but here the periodic coefficients are generated by the internal nonlinear driver.

To express these phases in frequency units, we denote the angular and cyclic frequencies of the driver by
\begin{equation}
    \omega_\kappa=\frac{2\pi}{T_\kappa},
    \qquad
    f_\kappa=\frac{1}{T_\kappa}.
    \label{eq:driver_frequencies}
\end{equation}
The lifted phase accumulated over one driver period then defines the angular Floquet quasifrequency
\begin{equation}
    \Omega_F(a):=
    \frac{\Theta_F(a)}{T_\kappa},
    \qquad
    r(a)=\frac{\Omega_F(a)}{\omega_\kappa}.
    \label{eq:lifted_frequency_ratio}
\end{equation}
The corresponding principal cyclic Floquet frequency, independent of the branch choice, is
\begin{equation}
    f_F:=
    \frac{\theta_F}{2\pi T_\kappa}
    =r_{\mathrm{pr}}f_\kappa.
    \label{eq:principal_floquet_frequency}
\end{equation}
These frequencies characterize the Floquet eigenphases. For Bloch-space observables, the relevant relative phase accumulates at twice the Floquet quasifrequency, reflecting the rotation angle $2\theta_F$ induced by each driver period. Moreover, the commensurability of this relative Floquet phase $2\theta_F$ with angle of $2\pi$ defines finite stroboscopic recurrence in the system.

\begin{figure}[t]
   \centering
   \includegraphics[width=8.0 cm]{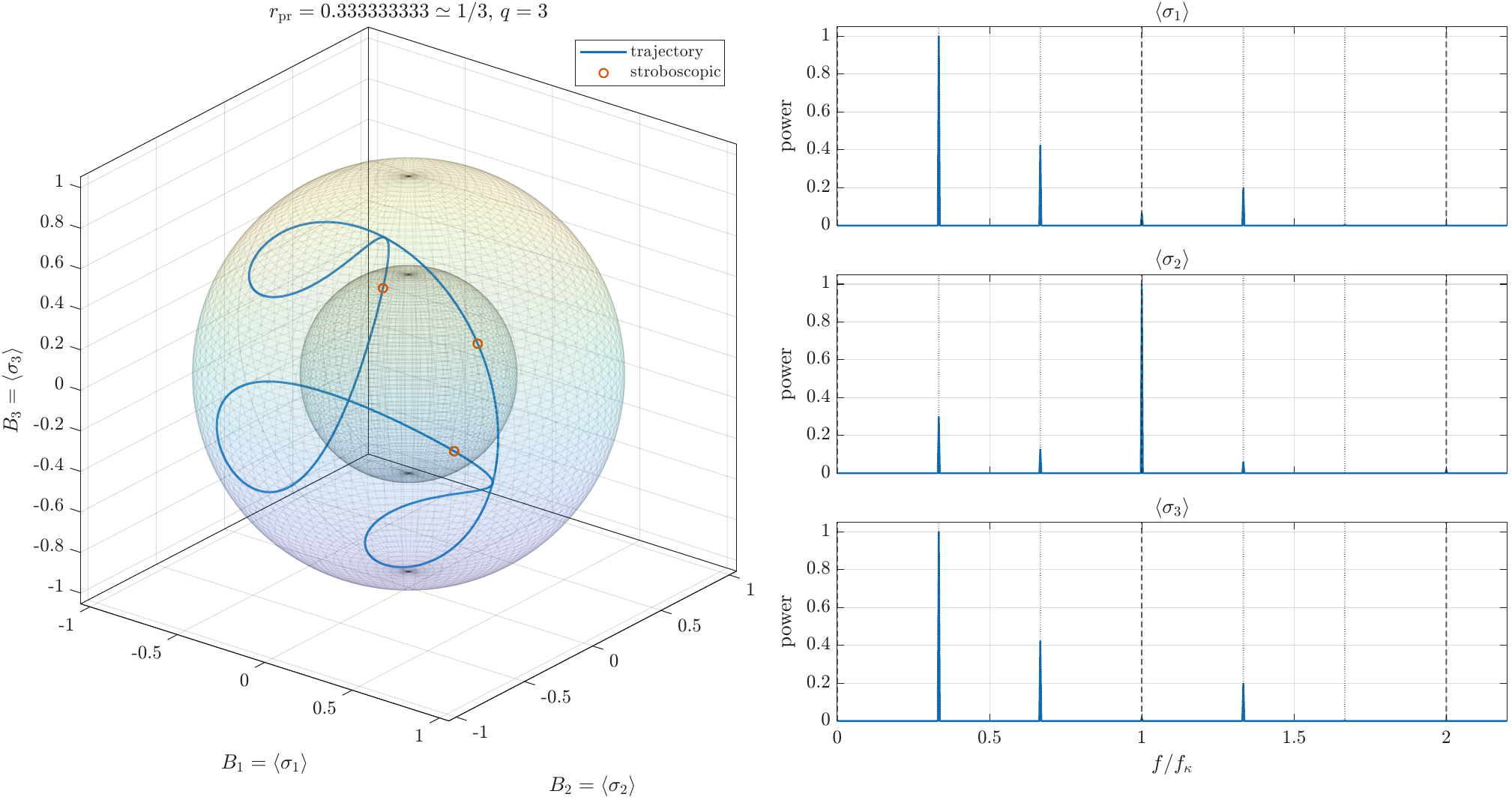}
   \caption{Rational stroboscopic recurrence for $r_{\mathrm{pr}}\simeq1/3$.
   Left: canonical Bloch trajectory $\mathbf B(t)$ in the Bloch space , with samples separated by $T_\kappa$ and closure after $q=3$ periods. Right: normalized spectra of the canonical components $B_i=\langle\sigma_i\rangle$; resolved peaks lie on $f/f_\kappa=\ell/3$.}
   \label{fig:rational_recurrence_main}
\end{figure}

Thus, one can show that for a rational principal ratio $r_{\mathrm{pr}}=j/n$, with $\gcd(j,n)=1$, the Bloch vector rotation in Eq.~\eqref{eq:bloch_rotation} closes after
\begin{equation}
    q=\frac{n}{\gcd(n,2)}
    \quad\text{driver periods},\qquad
    T_{\mathrm{obs}}=qT_\kappa.
    \label{eq:observable_recurrence}
\end{equation}
Indeed, $q$ applications of the evolution operator result in $(U_F)^q=\pm\mathbb{I}$, and the overall sign cancels for every quadratic observable. The coefficient matrix $Z(t)$ itself has the generic period of $nT_\kappa$. Generally, an exact recurrence of the original physical state vectors does not follow, as these may retain the external Hamiltonian evolution and the phase factors removed in defining $Z$. Further, for an irrational $r_{\mathrm{pr}}$, a generic Bloch trajectory does not close after any finite number of driver periods. Equivalently, its quadratic-observable spectrum contains sidebands
\begin{equation}
    f=kf_\kappa,
    \qquad f=kf_\kappa\pm 2r_{\mathrm{pr}}f_\kappa,
    \qquad k\in\mathbb Z,
    \label{eq:observable_sidebands}
\end{equation}
while a rational ratio generates the frequency lattice $f=\ell f_\kappa/q, \, l\in \mathbf{N}$. Shorter recurrence times may occur for specific initial states or exact observable projections.

To illustrate this point numerically, for the representative tuned case $p=-2$, $E\equiv N^2/V_{\min}=1.5$, $b=1$, $\lambda=0$, and $a=0.594543437307$, the numerical monodromy gives $r_{\mathrm{pr}}\simeq1/3$.  The corresponding three-period recurrence is shown in Fig.~\ref{fig:rational_recurrence_main} by the canonical Bloch trajectory and its stroboscopic samples; the associated spectral lattice confirms this independently.

The Floquet analysis, thus, identifies two complementary aspects of the nonlinear dynamics. The periodic $(\kappa,\tau)$ driver determines the instantaneous Bloch radius through the exact constraint $|\bm B|^2=1-ce^{p\kappa}/N^2$, while the monodromy $\mathcal{M}$ governs the stroboscopic reorientation of the Bloch trajectory. Consequently, the periodicity of the driver guarantees periodic radial motion, but does not generally imply the return of the two-level density matrix after a single driver period $T_\kappa$. Instead, successive periods are related by a fixed Bloch rotation, leading to finite stroboscopic recurrence when the Floquet ratio is rational and generically quasiperiodic motion otherwise. 

Note that the Bloch-space trajectories discussed here represent the exact evolution of the two-level representation of the density matrix under the nonlinear dynamics and they should be distinguished from the possible particle's trajectories in the physical space. The latter depends on the chosen observables and will be considered separately in Sec.~\ref{sec:real_trajectories}. 

Still, within the canonical Bloch representation, the Floquet eigenphase $\theta_F$ determines only the stroboscopic rotation angle $2\theta_F$ modulo $2\pi$, but not its axis. The latter, denoted by $\bm n_F^B$, provides the additional directional information needed to characterize the period-to-period reorientation of the trajectory. We therefore turn to the Floquet axis, its analytic small-oscillation approximation, and its numerical comparison with the full nonlinear dynamics.
\subsection{Floquet axes and their geometric representation}
\label{sec:floquet_axis}
The preceding analysis already established that boundary conditions provide our system with an explicit invariant Hermitian metric $\eta_G = G_0^{-1}$ for the monodromy as seen in Eq.~\eqref{eq:periodic_gram_floquet}. This invariant metric is, however, not unique. For comparison with the analytic small-oscillation calculation, we chose an eigenvector-based metric convention since it simplifies some of the calculations. To do that, away from the degenerate Floquet multipliers, we write the monodromy matrix through its diagonal representation as
\begin{equation}
    \mathcal M=S
    \begin{pmatrix}e^{i\theta_F}&0\\0&e^{-i\theta_F}\end{pmatrix}S^{-1},
\end{equation}
where the columns of $S$ are Euclidean-normalized eigenvectors. Then, one can define the following metric (that in general does not coincide with $\eta_G$):
\begin{equation}
    \eta_F=
    \frac{2(S^{-1})^\dagger S^{-1}}
    {\operatorname{Tr}[(S^{-1})^\dagger S^{-1}]},
    \qquad \operatorname{Tr}\eta_F=2.
    \label{eq:eta_convention}
\end{equation}
Choosing $L_F=\eta_F^{1/2}>0$, we introduce the $\eta_F$-adapted coordinates, $\widetilde{Z}=L_FZ$, so that the one period monodromy becomes an $SU(2)$ matrix admitting the usual decomposition:
\begin{equation}
    \widetilde{\mathcal M}_F=L_F\mathcal M L_F^{-1}\in SU(2),
    \qquad
    \widetilde{\mathcal M}_F
    =\cos\theta_F\,\mathbb{I}-i\sin\theta_F\,
      \mathbf n_F^{(\eta_F)}\cdot\boldsymbol\sigma.
    \label{eq:unitarized_etaF_axis}
\end{equation}
Away from the degenerate values of $\theta_F=0,\pi$, the unit vector $\mathbf n_F^{(\eta_F)}$ is the Floquet axis in this auxiliary $\eta_F$-frame. This axis serves as an intermediate quantity in the perturbative calculation. After an appropriate change of frame it can be identified with the canonical Bloch-space axis $\mathbf n_F^B$, defined by the physical stroboscopic density-matrix map in Eq.~\eqref{eq:bloch_rotation}. 

\subsection{Leading-order approximation for the Floquet axis}
We now derive the auxiliary Floquet axis in the small-oscillation regime, restricting to $\lambda=0$ as this parameter vanishes for quadratic observables. At zeroth order the $\kappa-\tau$ driver is at an equilibrium and the coefficient matrix $M$ is constant:
\begin{equation}
    M_0=\frac{N_0\mu}{2}\sigma_3-ie^{\kappa_0/2}\sqrt{a^2+b^2}\,\mathbf n_0\cdot\boldsymbol\sigma.
\end{equation}
where the auxiliary Pauli-vector coefficients are
\begin{equation}
    \mathbf n_0=\frac{1}{\sqrt{a^2+b^2}}
    \begin{pmatrix}
       a\cos\widetilde\theta_0+b\sin\widetilde\theta_0\\
       a\sin\widetilde\theta_0-b\cos\widetilde\theta_0\\0
    \end{pmatrix},
    \qquad
    \widetilde\theta_0=\frac{a\kappa_0}{2(b+\mu)}.
    \label{eq:nF0_explicit}
\end{equation}
The $\sigma_3$ term is Hermitian, while the transverse term is anti-Hermitian. The corresponding monodromy matrix is just
\begin{equation}
    \mathcal M^{(0)}=\exp(M_0T_0).
    \label{eq:M0_axis_form}
\end{equation}
Defining a new axis as $\mathbf m_0=\mathbf e_3\times\mathbf n_0$, we introduce the positive-definite change of frame
\begin{equation}
    L_0=\exp\!\left(
       \frac{\Lambda}{2}\mathbf m_0\cdot\boldsymbol\sigma
    \right),\qquad
    \Lambda=\operatorname{artanh}\!\left(\frac{N_0\mu/2}{e^{\kappa_0/2}\sqrt{a^2+b^2}}\right).
    \label{eq:L0_axis}
\end{equation}
For $e^{\kappa_0}(a^2+b^2)>(N_0\mu)^2/4$ this removes the Hermitian part of the generator and gives the zeroth order unitary monodromy:
\begin{equation}
    \widetilde{\mathcal M}^{(0)}=
    L_0\mathcal M^{(0)}L_0^{-1}
       =\exp\!\left(-i\omega_1T_0
          \mathbf n_0\cdot\boldsymbol\sigma\right),
    \qquad
    \omega_1^2=e^{\kappa_0}(a^2+b^2)
          -\left(\frac{N_0\mu}{2}\right)^2.
    \label{eq:M0_unitarized}
\end{equation}
Using $e^{\kappa_0}=-\mu N_0^2/(4b)$, this reproduces the frequency found in the analytic small-oscillation calculation (see Eq.~\eqref{eq:omega_1}).

Thus the leading-order Floquet axis in the $\eta_0$-frame is just $\bm n_0$ defined above. Using $R_3$ matrix of the ordinary $SO(3)$ rotation about the third axis that acts on the Pauli-vector coefficients in the $\eta_0$-frame, we write this result as
\begin{equation}
    \mathbf n_F^{(0,\eta_0)}\equiv \mathbf n_0
      =R_3(\widetilde\theta_0)
       \frac{1}{\sqrt{a^2+b^2}}
       \begin{pmatrix}a\\-b\\0\end{pmatrix},
    \qquad
    \widetilde\theta_0=\frac{a\kappa_0}{2(b+\mu)}.
    \label{eq:nF0_result}
\end{equation}
This axis is equatorial at zeroth order. For $b>0$ and $a=0$, it is $(0,-1,0)^T$ up to orientation. Switching on $a>0$ rotates it within the equatorial plane,
\begin{equation}
    \mathbf n_F^{(0,\eta_0)}(a)
       =R_3(\Delta\varphi_F^{(0)})
          \mathbf n_F^{(0,\eta_0)}(0),
    \qquad
    \Delta\varphi_F^{(0)}
       =\frac{a\kappa_0}{2(b+\mu)}+\arctan\frac ab.
    \label{eq:nF_rotation_from_a0}
\end{equation}

At nonzero driver amplitude $R$, the transformed monodromy $\widetilde{\mathcal M}$ in the fixed $\eta_0$ frame need not remain exactly unitary as we calculate the next perturbative order. The first order axis below refers to the monodromy's unitary (polar) part, as constructed in the Appendix~\ref{app:floquet_axis_correction}:
\begin{equation}
    \mathbf n_F^{(\eta_0)}
       =\mathbf n_F^{(0,\eta_0)}
        +R\,\delta\mathbf n_F^{(1,\eta_0)}+O(R^2),
    \label{eq:floquet_axis_expansion}
\end{equation}
where, away from the Floquet-zone edges,
\begin{equation}
    \delta\mathbf n_F^{(1,\eta_0)}
     =\frac{ae^{\kappa_0/2}\sqrt{a^2+b^2}}
            {(b+\mu)(\omega^2-4\omega_1^2)}
      \left[-2\omega_1\cos\theta_0\,\mathbf m_0
         +\omega\sin\theta_0\,\mathbf e_3\right],
    \qquad
    \mathbf m_0=\mathbf e_3\times
           \mathbf n_F^{(0,\eta_0)}.
    \label{eq:floquet_axis_first_order}
\end{equation}
The $\mathbf m_0$ component is an azimuthal correction, while the $\mathbf e_3$ component produces an out-of-plane tilt in the adapted representation. The correction is proportional to $a$ and vanishes for $a=0$ within this perturbative scheme.

\subsection{Numerical Floquet axes and comparison with the
small-oscillation approximation}
\begin{figure}[H]
\centering
    \includegraphics[width=8.0 cm]{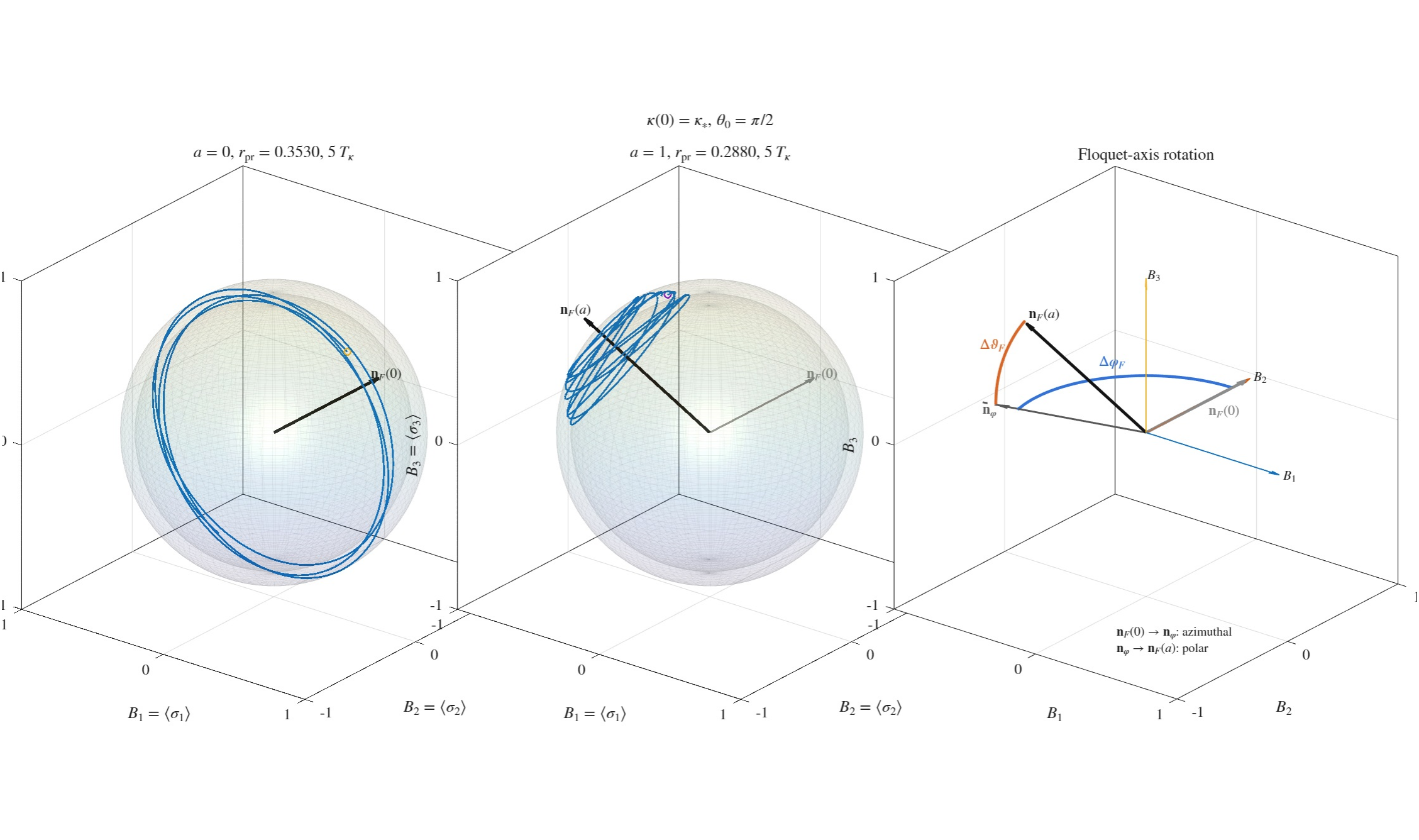}
    \caption{Bloch-space trajectories and Floquet-axis reorientation in the small-oscillation convention, for $p=-2$, $E=1.01$, $b=1$, $\lambda=0$, and $N=1$. The stroboscopic origin is chosen at $\kappa(0)=\kappa_0$, $\theta_0=\pi/2$, with $R=\sqrt{2(E-1)/|p|}=0.1$. Left and middle: five nonlinear-driver periods at $a=0$ and $a=1$, respectively. The driver and therefore the radial Bloch dynamics are unchanged, $|\mathbf B|_{\min}\simeq0.7735$ and  $|\mathbf B|_{\max}\simeq0.8548$, whereas the principal Floquet ratios change from $r_{\rm pr}\simeq0.3530$ to $r_{\rm pr}\simeq0.2880$. The black arrows indicate the canonical Bloch-space Floquet axes obtained from the exact unitary one-period map $U_F$, independently validated against the stroboscopic trajectory fit; for $a=0$, $\mathbf n_F^B\simeq(0,1,0)$, and for $a=1$ the displayed representative is $\mathbf n_F^B\simeq(-0.7877,-0.1889,0.5864)$. Right: decomposition of this displayed reorientation into $\Delta\varphi_F\simeq103.49^\circ$ and $\Delta\vartheta_F\simeq-35.90^\circ$. These angles are defined in the canonical Bloch-space.}
    \label{fig:traj-nF}
\end{figure}

For the numerical analysis it is essential to distinguish the axis of the canonical Bloch-space rotation (denoted by '$B$') from the axes in the auxiliary frame ('$(\eta_F)$' is numerically calculated from the monodromy diagonalization and '$(j)$' denotes the perturbative order). The former follows directly from the one period evolution operator $U_F$ established above. Thus, the one-period rotation of the Bloch vector $B$ is defined by
\begin{equation}
    \bigl(Q_F^B\bigr)_{ij}
    =\frac12\operatorname{Tr}\!\left(
        \sigma_i U_F^T\sigma_j U_F^*
    \right),
    \qquad
    Q_F^B\mathbf n_F^B=\mathbf n_F^B,
    \label{eq:canonical_Q_from_U}
\end{equation}
where $\bm n^B_F$ is a Floquet axis in the canonical Bloch space. For the corresponding Bloch trajectories in Fig.~\ref{fig:traj-nF}, the canonical axes at $a=0$ and $a=1$ have an angular separation defined by
\begin{equation}
    \Delta_B
    =\arccos\!\left|
        \mathbf n_F^B(0)\cdot\mathbf n_F^B(1)
    \right|
    \simeq79.11^\circ.
    \label{eq:canonical_axis_separation}
\end{equation}
The exact construction agrees with the independent stroboscopic trajectory fit to numerical precision.

To assess the small-oscillation approximation, we compare the analytic and exact Floquet axes in the common $\eta_F$-frame, using the convention of Eq.~\eqref{eq:eta_convention}. The approximate axes are extracted from the unitary polar factors of the transformed monodromies. Their angular errors in the auxiliary $\eta_F$-frame are
\begin{equation}
\begin{aligned}
    \Delta_{\mathrm{axis}}^{(j)}
    &=\arccos\!\left|
        \mathbf n_F^{((j),\eta_F)}
        \cdot\mathbf n_F^{(\eta_F,\mathrm{exact})}
    \right|,
    \qquad j=0,1,\\
    \Delta_{\mathrm{axis}}^{(0)}
    &\simeq14.52^\circ,\qquad
    \Delta_{\mathrm{axis}}^{(1)}
    \simeq2.29^\circ
    \quad (a=1,\ R=0.1).
\end{aligned}
\label{eq:axis_error_metric_frame}
\end{equation}

%%%%%%%%%%%%%%%%%%%%%%%%%%%%%%%%%%%%%
\section{Analytical and numerical results on singular limits and observables}
\label{sec:numeric_res}
Unless stated otherwise, for each parameter point we first integrate the nonlinear $(\kappa,\tau)$ subsystem and determine the driver period $T_\kappa$ from the corresponding return event.  We then integrate the fundamental matrix over one period, diagonalize the monodromy matrix $\mathcal M=\mathcal Z(T_\kappa)$, and extract its Floquet multipliers, phase $\theta_F$, and principal ratio $r_{\mathrm{pr}}$.  Finally, we integrate the state-vector dynamics and reconstruct the two-level density matrix and the associated Bloch trajectory.  For the numerical diagnostics we monitor the state normalization and boundary constraints as defined in Eq.~\eqref{eq:boundary_conditions}, the monodromy determinant of one and a unit-circle error, and the recurrence errors of the state and selected observables.

The analytic discussion above expressed expectation values in terms of the vector $\vec u$ in Eq.~\eqref{eq:ZZ_dagger_in_u} and Eq.~\eqref{eq:vec_u_squared}). For the numerical trajectories it is convenient to use instead the canonical Bloch vector $\mathbf B$ of the two-level density matrix in the fixed $\{|A_i\rangle\}$ basis as defined in the previous section (see Eq.~\eqref{eq:bloch_prior_to_recurrence}). The trajectories shown below, therefore, are plotted in the canonical Bloch coordinates unless stated otherwise. 

\subsection{Small-oscillation approximation: analytic versus numerical solutions}
\label{sec:small_osc_numerics}

We now compare the first-order small-oscillation solution derived in Sec.~\ref{sec:small_osc_analytic} with the full numerical dynamics. The nonlinear driver expanded about the equilibrium is harmonic to the leading order:
\begin{equation}
    \kappa(t)
    =
    \kappa_0
    +
    R\cos(\omega t-\theta_0)
    +
    O(R^2),
\end{equation}
together with the corresponding first-order expression for $\tau(t)$. As obtained in Sec.~\ref{sec:small_osc_analytic}, the state vectors matrix solution can be written in the following form:
\begin{equation}
    Z_{\mathrm{an}}(t)
    =
    e^{M_0t}
    \left[
        \mathbb{I}
        +
        R\left(W_1+F(t)\right)
    \right]
    Z_0(0)
    +
    O(R^2),
    \label{eq:small_osc_Z_an}
\end{equation}
where $F(t)$ is the interaction-picture correction and the constant matrix $W_1$ is fixed by the first-order matching of the initial data. The corresponding first-order fundamental matrix over the small-oscillation period, $T_0=\frac{2\pi}{\omega}$, is
\begin{equation}
    \mathcal{Z}_{\mathrm{an}}(T_0)
    =
    e^{M_0T_0}
    \left[
        \mathbb{I}+R F(T_0)
    \right]
    +
    O(R^2),
    \label{eq:small_osc_monodromy_an}
\end{equation}
which also defines the first-order monodromy $\mathcal M_{\mathrm{an}}$. The correction to the exact nonlinear driver period, $T_\kappa(R)-T_0$, is itself of order $R^2$ and therefore does not enter the first-order approximation.

For the numerical validation, the full nonlinear $(\kappa,\tau)$ system is integrated without linearization.  The resulting numerical $\kappa(t)$ is then used in the full time-dependent $2\times2$ coefficient matrix $M(t)$, from which both $Z(t)$ and the fundamental matrix $\mathcal{Z}(t)$ are obtained numerically.  Analytic and numerical solutions are initialized with the same initial data.  To test the local asymptotic accuracy independently of long-time phase accumulation, the state-space errors are evaluated over a fixed interval of one small-oscillation period, $0\leq t\leq T_0$.

We quantify the fixed-time errors by\footnote{$F$ stands for the Frobenius norm or the Hilbert–Schmidt norm.}
\begin{align}
    \epsilon_Z(R)
    =
    \max_{0\leq t\leq T_0}
    \left\|
        Z_{\mathrm{num}}(t)-Z_{\mathrm{an}}(t)
    \right\|_F, \qquad
    \epsilon_B(R)
    =
    \max_{0\leq t\leq T_0}
    \left|
        \bm B_{\mathrm{num}}(t)-\bm B_{\mathrm{an}}(t)
    \right|,
\end{align}
and analogously for the selected observables.  For the Floquet data we compare the exact one-period monodromy with the first-order expression, and define its relative error together with the principal Floquet-ratio error as
\begin{equation}
    \epsilon_\Phi(R)
    =
    \frac{
        \left\|
            \mathcal{M}_{\mathrm{num}}(T_\kappa)
            -
            \mathcal{M}_{\mathrm{an}}(T_0)
        \right\|_F
    }{
        \left\|
            \mathcal{M}_{\mathrm{num}}(T_\kappa)
        \right\|_F
    }, \quad
    \epsilon_r(R)
    =
    \left|
        r_{\mathrm{num}}-r_{\mathrm{an}}
    \right|.
\end{equation}
Since the analytic construction retains all terms through first order in $R$, all of these quantities are expected to have a quadratic remainder, $\epsilon(R)=O(R^2)$.

\begin{figure}[H]
    \centering
    \includegraphics[width=8.0 cm]{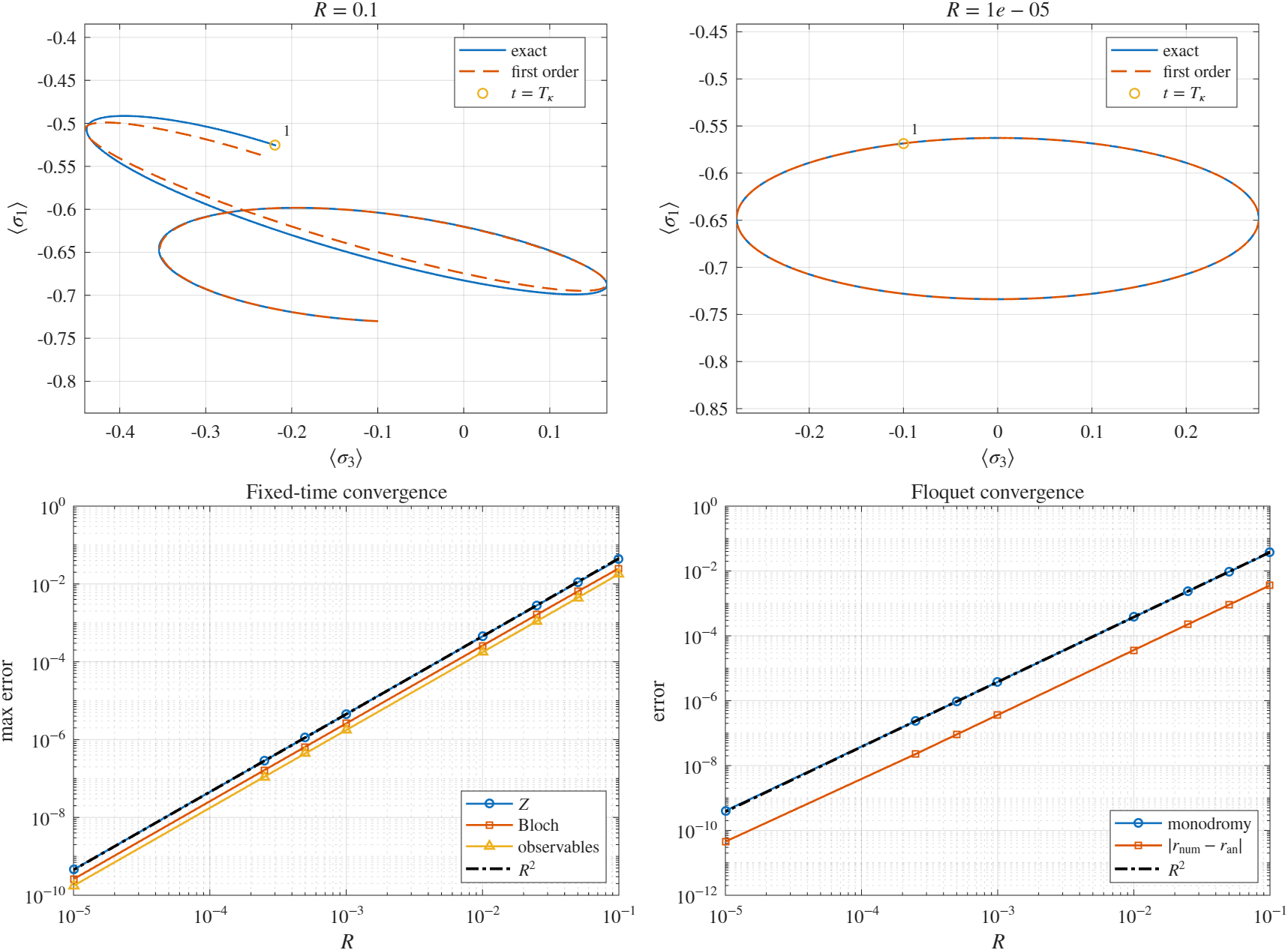}
    \caption{Validation of the first-order small-oscillation approximation for $p=-2$, $a=1$, $b=1$, $\lambda=0$, and $N_0=1$. The upper panels compare the numerical Bloch-space trajectory $(\langle\sigma_3\rangle,\langle\sigma_1\rangle)$ with the first-order analytic solution for $R=0.1$ (left) and $R=10^{-5}$ (right). The marked point denotes the end of one exact nonlinear driver period, $t=T_\kappa$.  At $R=0.1$ finite-amplitude corrections are visible, whereas at $R=10^{-5}$ the two solutions are indistinguishable at the scale of the plot. The lower-left panel shows the maximum fixed-time errors in $Z$, the Bloch vector, and the selected observables over one small-oscillation period, $T_0$. The lower-right panel shows the relative monodromy error and the principal Floquet-ratio error $|r_{\mathrm{num}}-r_{\mathrm{an}}|$. The dash-dotted reference lines indicate the predicted $R^2$ scaling, which is reproduced numerically over the displayed amplitude range.}
    \label{fig:small_osc_val}
\end{figure}

Figure~\ref{fig:small_osc_val} confirms this behavior over several orders of magnitude in $R$. For the representative parameters $p=-2$, $a=1$, $b=1$, $\lambda=0$, and $N_0=1$, log-log fits give slopes of approximately $2$ ($1.996,~ 1.997,~1.998$ for the errors in $Z$, the Bloch vector, and the plotted observables, respectively). The corresponding convergence exponents for the relative monodromy error and for $|r_{\mathrm{num}}-r_{\mathrm{an}}|$ have the same expected behavior (approximately $1.99$ and $1.98$). The upper panels of Fig.~\ref{fig:small_osc_val} illustrate the same result using Bloch-space trajectories.  

\begin{figure}[H]
\centering
\includegraphics[width=8.0 cm]{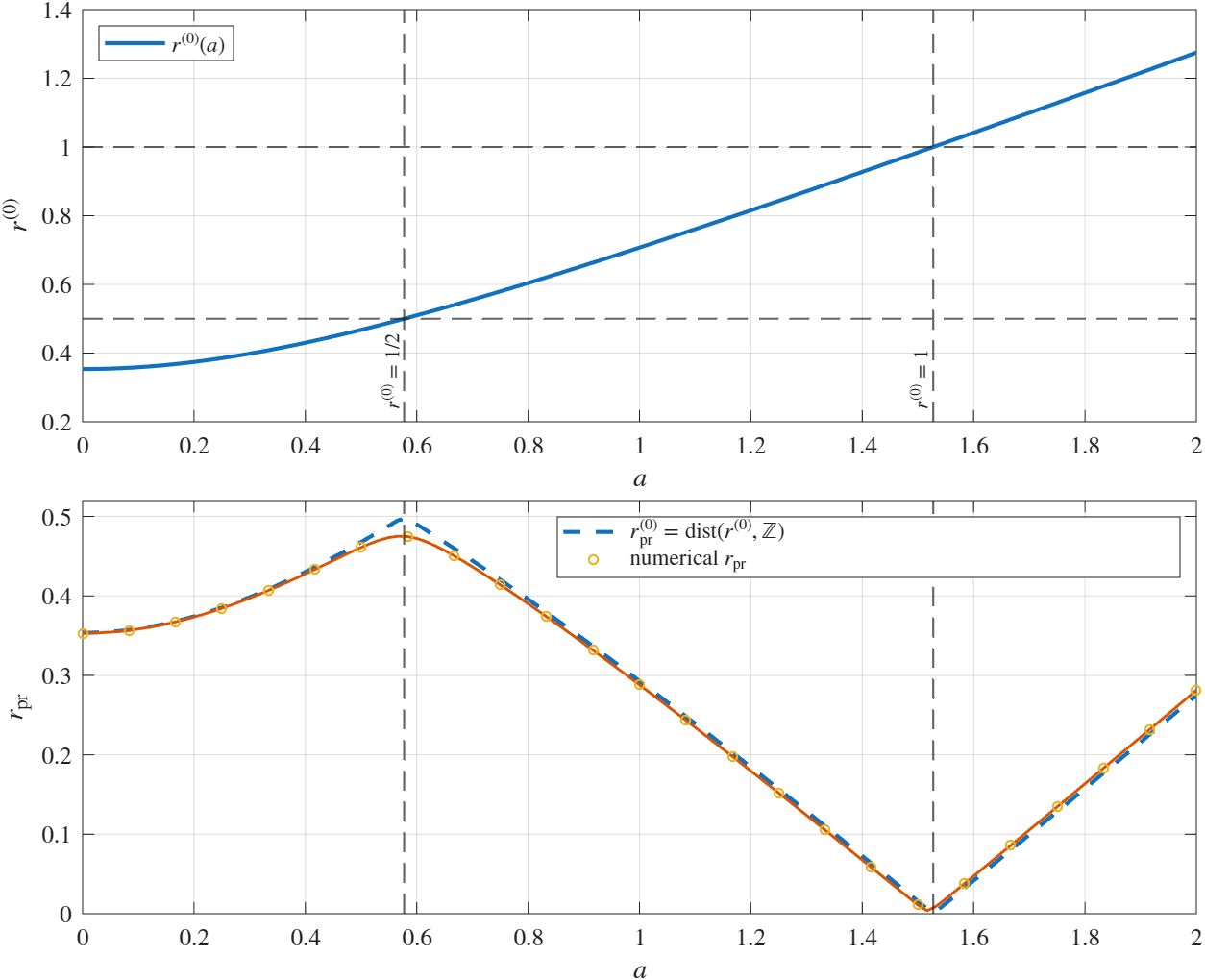}
\caption{Floquet-zone folding as $a$ is varied, for $p=-2$, $b=1$, $\lambda=0$, $N=1$, and $E=1.01$. The upper panel shows the lifted small-oscillation prediction $r^{(0)}(a)$, while the lower panel compares its principal-zone folding, $r_{\mathrm{pr}}^{(0)}=\operatorname{dist}(r^{(0)},\mathbb{Z})$, with the principal Floquet ratio obtained numerically from the one-period monodromy matrix.  The close agreement shows that the repeated branches of $r_{\mathrm{pr}}$ arise from Floquet-zone folding of a smoothly increasing lifted frequency.}
  \label{fig:a_folding}
\end{figure}

In the small-oscillation regime, the analytic approximation provides a useful framework to illustrate the effect of the Floquet zone folding discussed in the previous section. Here we have a natural choice of the parameter branch for the lifted Floquet ratio (or frequency) as defined in Eq. (\ref{eq:lifted_floquet_ratio}). We denote the small-oscillation approximation of it by $r^{(0)}(a)$ and choose the branch continuously from $a=0$. Using the formulas for $\omega_1 \equiv \Omega_F$ and $\omega \equiv \omega_\kappa$ from the analytic section (see Eq.~\eqref{eq:omega'},\eqref{eq:tau_approx}, \eqref{eq:omega_1}) as the approximations for the Floquet frequency and the driver frequency accordingly, at $\lambda=0$ and $b=1$ one obtains:
\begin{equation}
    \left[r^{(0)}(a)\right]^2
    \simeq
    \frac{1+(1+|p|)a^2}{8}.
    \label{eq:small_oscillation_lifted_ratio}
\end{equation}
The corresponding principal-zone prediction is, therefore, defined by
\begin{equation}
    r_{\mathrm{pr}}^{(0)}(a)
    =
    \operatorname{dist}
    \bigl(r^{(0)}(a),\mathbb{Z}\bigr).
    \label{eq:small_oscillation_principal_ratio}
\end{equation}
At $a=0$ the small oscillations approximation has its minimum of $r^{(0)} \simeq \frac{1}{2\sqrt{2}}$. While this value can be interpreted as the baseline of the lifted parameter branch, it is not the lower bound on the principal ratio $r_{pr}$ itself, which can approach zero after the Floquet-zone folding. For $|p|=2$, for example, we get
\begin{equation}
    \left(r^{(0)}\right)^2 \simeq \frac{1+3a^2}{8},
\end{equation}
with the first principal maximum and first fold occurring approximately at
\begin{equation}
    r^{(0)} \simeq \frac12~\text{and}~ r_{pr}^{(0)} \simeq\frac{1}{2}
    \text{ at }
    a=\frac{1}{\sqrt{3}}, \qquad 
    r^{(0)}\simeq1~\text{and}~ r^{(0)}_{pr} \simeq0
    \text{ at }
    a=\sqrt{\frac{7}{3}}.
\end{equation}
These folds are also shown on Figure \ref{fig:a_folding}. This way, whenever $r$ passes through an integer, the principal ratio returns to zero; whenever $r$ passes through a half-integer, it reaches the edge of the principal Floquet zone (fundamental zone), $r_{\rm pr}=\frac12$. The apparent repetition of ratio bands is a zone-folding effect, analogous to folding Bloch branches into a fundamental quasienergy or Brillouin zone. 

\subsection{The small-$c$ regime}
\label{sec:smallc-singular}
The analytic $c=0$ solution provides an exact ordinary quantum mechanical boundary of the model.  We now ask how the neighboring periodic $c>0$ solutions approach this limit.  The solutions family is obtained by keeping the equations of motion fixed at
\begin{equation}
    a,\ b,\ \mu,\ \lambda,\ N=\mathrm{const},
    \qquad
    c\to0^+ .
    \label{eq:smallc-fixed-couplings}
\end{equation}
Thus, $c$ varies only through the initial configuration, while these combinations
\begin{equation}
    \alpha\equiv b+\mu,
    \qquad
    p=\frac{\mu}{b+\mu}<0
\end{equation}
remain fixed.

For $p<0$ the driver potential $V(\kappa)=4e^\kappa+ce^{p\kappa}$ has a unique minimum,
\begin{equation}
    V_{\min}
    =
    (1-p)
    \left(-\frac{p}{4}\right)^{\frac{p}{1-p}}
    c^{\frac{1}{1-p}}\quad \text{ at }
    \quad
    \kappa_0
    =
    \frac{1}{1-p}
    \log\!\left(-\frac{pc}{4}\right).
    \label{eq:smallc-Vmin}
\end{equation}
It is therefore convenient to introduce the dimensionless energy by normalizing the first integral $N^2$ as
\begin{equation}
    E(c)
    \equiv
    \frac{N^2}{V_{\min}}
    =
    \frac{N^2}
    {(1-p)\left(-p/4\right)^{p/(1-p)}}
    c^{-1/(1-p)} .
    \label{eq:smallc-E}
\end{equation}
Note, that along the parameter family \eqref{eq:smallc-fixed-couplings}, one can see the equivalence of the two limits:
\begin{equation}
    c\to0^+
    \qquad\Longleftrightarrow\qquad
    E(c)\to\infty .
\end{equation}
In what follows, this large $E$ limit is used as a convenient asymptotic parameter for the fixed-$N$, fixed-$p$ small-$c$ path.

To study the dynamics of the proposed limit, we first note that the Bloch radius $R_B$ is fixed directly by the $\kappa-\tau$ driver,
\begin{equation}
    R_B^2 \equiv |\vec B|^2 \
    =
    1-\frac{ce^{p\kappa}}{N^2}
    =
    \frac{\tau^2+4e^\kappa}{N^2}.
    \label{eq:smallc-RB}
\end{equation}
For small $c$, the two turning points, $\kappa_\pm$, separate strongly, since the left side of the potential $V_\kappa$ becomes increasingly flat. Combining this with the dependency of the Bloch radius on $\kappa$, at fixed $N$ and $p$ one finds the corresponding limits for the Bloch radii:
\begin{equation}
    R_{B,\min}^2=O\!\left(c^{1/|p|}\right)\to0,
    \qquad
    1-R_{B,\max}^2=O(c)\to0.
    \label{eq:smallc-radial-range}
\end{equation}
Thus, a finite-$c$ orbit develops an almost complete radial excursion. The Bloch radius changes over the entire $(0,1)$ interval, even though its $c=0$ limit lies entirely on the pure-state boundary $R_B=1$. Moreover, the deep inward excursion is pushed to progressively later times, occupying an asymptotically vanishing fraction of the full driver period, as we shall show below.

\begin{figure}[H]
\centering
    \includegraphics[width=8.0 cm]{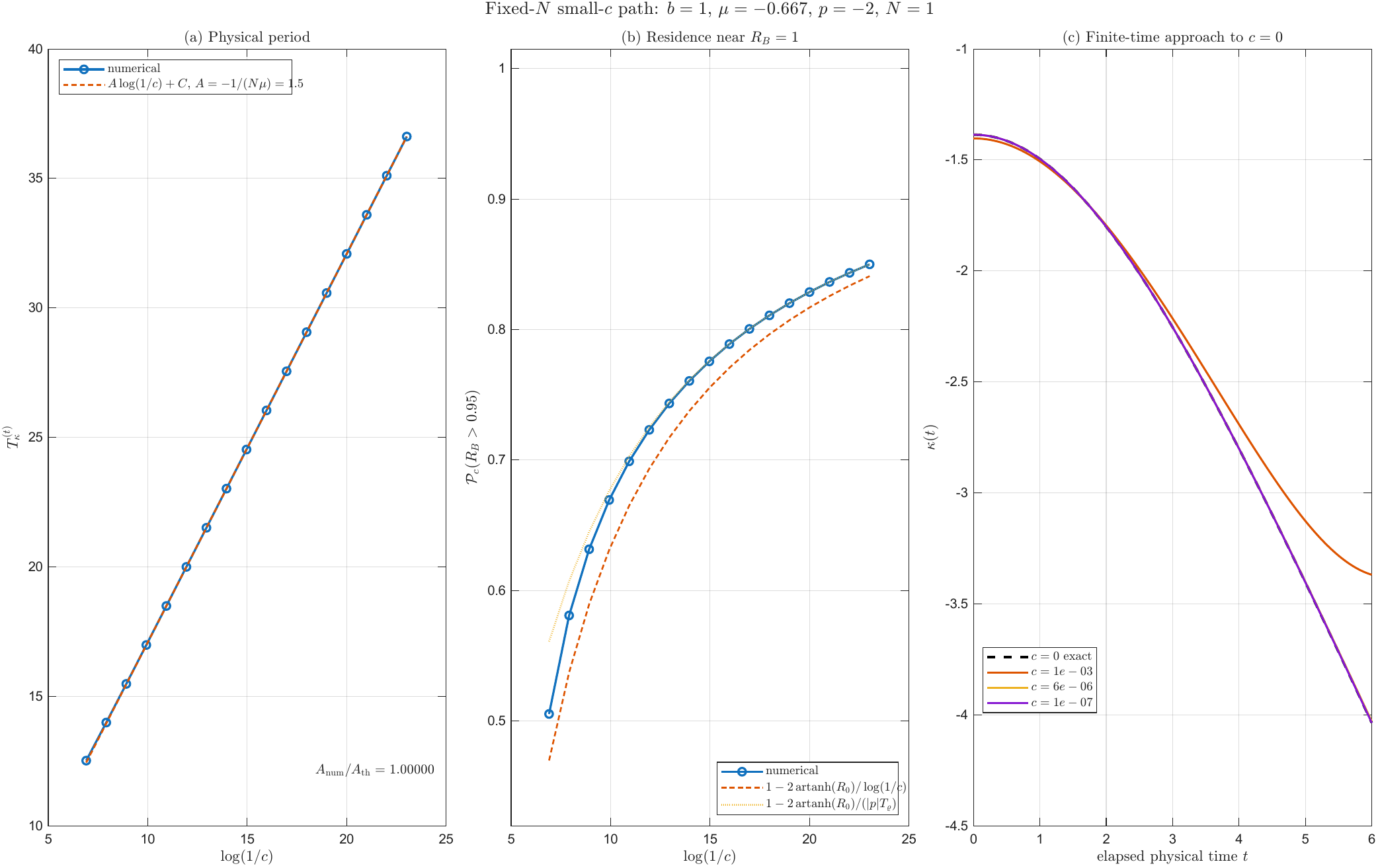}
    \caption{Approach to the $c=0$ ordinary-QM boundary along the fixed-$p$ path, i.e. $c\to0^+$ with $a,b,\mu,\lambda$, and $N$ held fixed. (a) The physical driver period grows logarithmically, $T_{\kappa}^{(t)}\sim -(N\mu)^{-1}\log(1/c)$; the numerical slope agrees with the asymptotic prediction. (b) The fraction of one period spent near the pure-state Bloch boundary, $\mathcal P_c(R_B>R_0)$, increases toward unity as $c\to0$.  The dashed curve shows the leading small-$c$ form $1-2\,\mathrm{artanh}(R_0)/\log(1/c)$, while the dotted curve shows the same prediction but with the numerically determined normalized period $T_{\varrho}$. (c) On any fixed physical-time interval, the finite-$c$ driver converges to the exact $c=0$ solution. On a larger scale, the finite-$c$ solutions remain periodic, while their return time diverges logarithmically.}
    \label{fig:smallc-fixed-couplings-summary}
\end{figure}

To quantify this effect, we introduce the following short-hands for the centered/normalized $\kappa-\tau$ pair and a normalized time:
\begin{equation}
    \hat\kappa=\kappa-\kappa_0,
    \qquad
    \hat\tau=\frac{\tau}{N},
    \qquad
    \varrho=-\alpha Nt .
\end{equation}
For every fixed arbitrary radius $R_0<1$, the (normalized) time spent in the interior region of the Bloch ball, $R_B\le R_0$, has the large-$E$ limit of
\begin{equation}
    \Delta\varrho(R_B\le R_0)
    =
    \frac{2}{-p}\operatorname{artanh}R_0+o(1),
    \label{eq:smallc-interior-rho}
\end{equation}
whereas the full normalized period grows as
\begin{equation}
    T_\varrho(E)
    =
    \left(1-\frac{1}{p}\right)\log E+O(1)=
    -\frac{1}{p}\log\frac{1}{c}+O(1).
    \label{eq:smallc-Trho}
\end{equation}
We define the near-boundary residence probability as the fraction of one driver period spent at $R_B>R_0$. Since normalized time $\varrho$ is affine in physical time, this fraction is independent of the time parametrization:
\[
    \mathcal P_c(R_B>R_0)
    =
    1-\frac{\Delta\varrho(R_B\le R_0)}
            {T_\varrho(E)}.
\]
Combining the two asymptotic expressions gives 
\begin{equation}
    \mathcal P_c(R_B>R_0)
    =
    1-
    \frac{2\operatorname{artanh}R_0}
         {\log(1/c)}
    +o\!\left((\log(1/c))^{-1}\right).
    \label{eq:smallc-residence}
\end{equation}
Similarly, for the impurity one can write
\begin{equation}
    1-\left\langle\operatorname{Tr}\rho^2\right\rangle
    =
    \frac{1}{\log(1/c)}
    +o\!\left((\log(1/c))^{-1}\right).
    \label{eq:smallc-impurity}
\end{equation}
Hence, the system indeed spends an asymptotically dominant fraction of each finite-$c$ period close to the pure-state boundary as $c\to0^+$. The predicted logarithmic divergence of the driver period and the increasing near-boundary residence probability are confirmed numerically in Fig.~\ref{fig:smallc-fixed-couplings-summary}(a,b). In particular, panel (b) shows the agreement with the leading asymptotic expression \eqref{eq:smallc-residence}, as well as the improved estimate obtained by retaining the numerically evaluated period $T_\varrho(E)$.

The asymptotics above are written in terms of the normalized time $\varrho$. Accordingly, the physical period follows from $\varrho=-(b+\mu) Nt$ and Eqs.~\eqref{eq:smallc-E}-\eqref{eq:smallc-Trho}:
\begin{equation}
    T_\kappa^{(t)}
    \sim
    \frac{1}{N(b-\alpha)}
    \log\frac{1}{c}
    =
    -\frac{1}{N\mu}
    \log\frac{1}{c} \rightarrow \infty \quad\text{as} \quad c\rightarrow 0^+.
    \label{eq:smallc-Tphysical}
\end{equation}
Note, that for the $p<0$ branch considered here, $\mu<0$, so the coefficient is positive. By contrast, the absolute physical time spent below a fixed radius $R_0<1$ remains finite:
\begin{equation}
    \Delta t(R_B\le R_0)
    \longrightarrow
    \frac{2\,\operatorname{artanh}R_0}
         {(-\mu)N}.
    \label{eq:smallc-interior-t}
\end{equation}
As illustrated in Fig.~\ref{fig:smallc-fixed-couplings-summary}(c), the finite-$c$ solutions approach the exact $c=0$ dynamics on fixed physical-time intervals, even though their return period diverges. The small-$c$ limit is regular only on fixed physical-time intervals. The finite-$c$ solution follows the exact $c=0$ trajectory for an increasingly long time, until the rapid excursion closes the $c>0$ orbit at times of order $\log(1/c)$.  This is the finite-$c$ realization of the ordinary-QM boundary found analytically at exact $c=0$. 

\subsection{Large $|p|$ and the approach to $b+\mu=0$}
\label{sec:smallc-largep}
The fixed-coupling small-$c$ limit above keeps $p$-parameter finite and fixed.  A distinct singular direction is obtained by increasing $|p|$.  Since
\begin{equation}
    p
    =
    \frac{\mu}{b+\mu}
    =
    1-\frac{b}{\alpha},
    \qquad
    \alpha\equiv b+\mu,
    \label{eq:largep-palpha}
\end{equation}
one has at fixed finite $b>0$ an equivalence of the following two limits:
\begin{equation}
    p\to-\infty
    \quad\Longleftrightarrow\quad
    \alpha\to0^+,
    \qquad
    \mu\to-b.
    \label{eq:largep-alpha0}
\end{equation}
Large negative $p$, therefore, represents an approach to the analytically distinguished surface $b+\mu=0$.

\begin{figure}[t]
    \centering
    \includegraphics[width=8.0 cm]{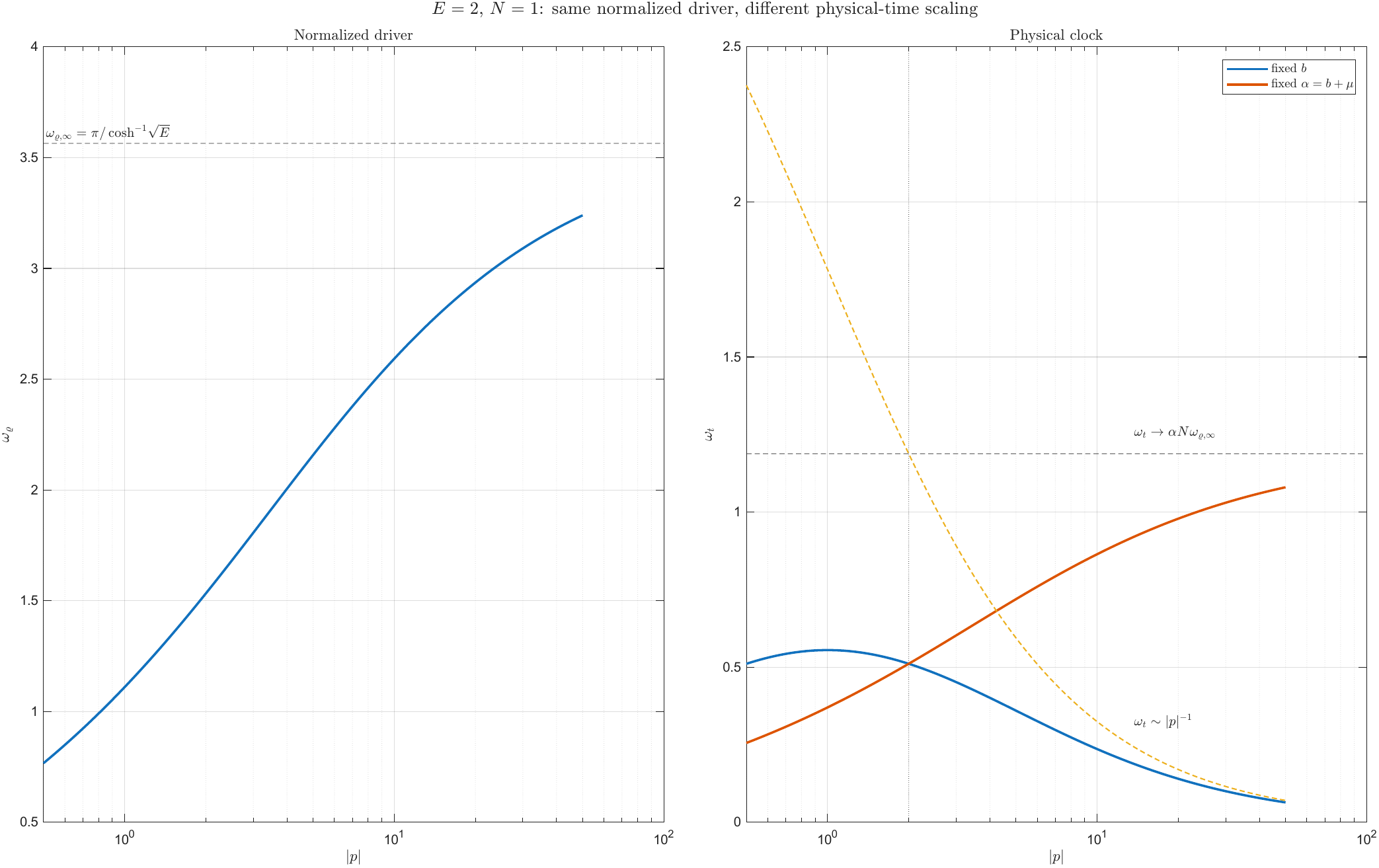}
    \caption{Separation of normalized driver dynamics from physical-time scaling in the large-$|p|$ regime, shown at fixed $E=2$ and $N=1$.  For each $|p|$, the fixed-$b$ and fixed-$\alpha$ families correspond to the same normalized driver and therefore share the same $\omega_{\varrho}$, which approaches the finite limit $\omega_{\varrho,\infty}=\pi/\cosh^{-1}\!\sqrt{E}$ (left).  Their physical frequencies differ through $\omega_t=\alpha N\omega_{\varrho}$ (right).  At fixed $b$, $\alpha=b/(1+|p|)\to0$, giving $\omega_t\sim |p|^{-1}\to0$, whereas at fixed $\alpha=b+\mu$ the physical frequency approaches the finite value $\alpha N\omega_{\varrho,\infty}$.  Thus the physical slowdown at large $|p|$ is caused by the simultaneous approach to the singular surface $b+\mu=0$, rather than by large $|p|$ itself.}
    \label{fig:largep-clock-control}
\end{figure}

In this regime it is especially important to distinguish between the normalized time $\varrho$ and physical time $t$. The corresponding frequencies are related by
\begin{equation}
    \omega_t
    =
    \alpha N\,\omega_\varrho .
    \label{eq:largep-clock}
\end{equation}
Hence, a driver that remains finite-frequency in normalized $\varrho$-time can become arbitrarily slow in physical time because $\alpha=b+\mu$ tends to zero. A useful control here is to vary $|p|$ while holding the coefficient $\alpha=b+\mu$ fixed (see Fig. \ref{fig:largep-clock-control}).  Then,
\begin{equation}
    b=(1+|p|)\alpha,
    \qquad
    \mu=-|p|\alpha,
\end{equation}
so the physical clock factor remains fixed while the parameters $b$ and $\mu$ grow in magnitude.

\begin{figure}[H]
\centering
    \includegraphics[width=10.0 cm]{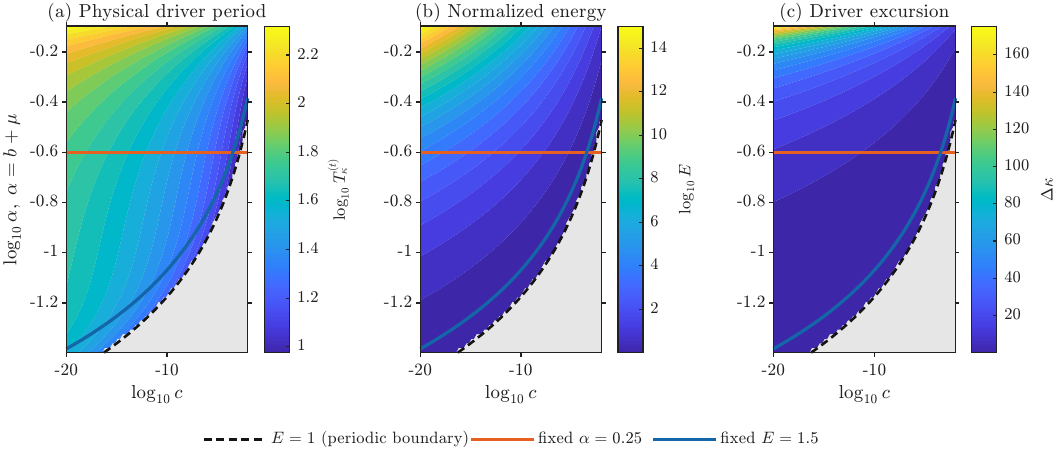}
    \caption{Structure of the periodic sector near the two singular boundaries $c\to0^+$ and $\alpha=b+\mu\to0^+$, for $b=N=1$. The parameter maps show (a) the logarithm of the physical driver period $\log_{10}T_\kappa^{(t)}$, (b) the normalized energy $\log_{10}E$, where $E=N^2/V_{\min}$, and (c) the driver excursion $\Delta\kappa=\kappa_+-\kappa_-$. The dashed black curve marks the boundary $E=1$, separating periodic motion ($E>1$) from the nonperiodic region (gray). The horizontal line represents the fixed-$\alpha=0.25$ path toward the ordinary-quantum-mechanical limit $c\to0^+$, while the curved line follows $E=1.5$ toward the exceptional boundary $\alpha\to0^+$.}
    \label{fig:smallc-alpha-corner}
\end{figure}

The two singular limits discussed meet at the corner
\begin{equation}
    (c,\alpha)=(0,0).
\end{equation}
The structure of the periodic sector near this corner is illustrated in Fig.~\ref{fig:smallc-alpha-corner}. For the chosen $N=b=1$ family, the periodic region is characterized by condition $E>1$; it is bounded by the contour $E=1$, shown as the dashed black curve in panel (b). Consequently, a path with fixed $c>0$ and decreasing $\alpha$ crosses this boundary before reaching the exact limit $\alpha=0$. A periodic approach to the $b+\mu=0$ boundary, therefore, requires $c$ to decrease simultaneously.

The two paths highlighted in Fig.~\ref{fig:smallc-alpha-corner} illustrate this effect. The horizontal path at fixed $\alpha=0.25$ approaches the ordinary-QM boundary $c\to0^+$ remaining in the periodic sector. By contrast, the curved path keeps
\begin{equation}
    E=E_*>1
\end{equation}
fixed and approaches $\alpha=0^+$ without crossing the periodicity boundary. Writing $|p|=b/\alpha-1$, this correlated path is given by
\begin{equation}
    c_{E_*}(|p|)
    =
    \left[
        \frac{N^2}{E_*(1+|p|)}
    \right]^{1+|p|}
    \left(\frac{|p|}{4}\right)^{|p|}.
    \label{eq:largep-fixedE-path}
\end{equation}
As $p$ and $E$ are not properly defined at the exact limit of $\alpha = 0$, for $a=\lambda=0$, the fixed-$E_*$ path selects a particular solution of the limiting $b+\mu=0$ dynamics, with
\begin{equation}
    2\omega_0
    =
    N\sqrt{1-\frac{1}{E_*}},
    \qquad
    e^\kappa
    =
    \frac{N^2}{4E_*},
\end{equation}
and
\begin{equation}
    \tau(t)
    =-2\omega_0\tanh(2\omega_0bt),
    \label{eq:largep-alpha0-solution}
\end{equation}
in agreement with Eq.~\eqref{eq:tau_b_and_mu_0}.

Thus, the joint $(c,\alpha)\to(0,0)$ limit is strongly path dependent. The fixed-$\alpha$ $c\to0^+$ path approaches the ordinary-QM boundary, whereas the correlated fixed-$E_*>1$ path selects a nontrivial member of the $b+\mu=0$ solution family. The accompanying changes in the driver period and excursion are displayed in Fig.~\ref{fig:smallc-alpha-corner}(a,c), respectively, emphasizing that the two paths approach the same parameter-space corner through dynamically distinct regimes.

\subsection{From Bloch-space motion to physical position trajectories}
\label{sec:real_trajectories}
The preceding analysis describes the dynamics in terms of the canonical Bloch vector $\mathbf B(t)$ associated with the two-level density matrix. This representation is particularly natural for  the Floquet analysis, since the stroboscopic map acts as a rotation in Bloch space. However, the Bloch vector is not itself a particle trajectory in a physical space. To obtain a trajectory of position expectation values, one must specify the position observables and a corresponding linear map:
\begin{equation}
    r_k(t):=\langle X_k'\rangle(t)
    =(C\mathbf B(t))_k,
    \qquad
    C_{kj}=(\vec X_k^T J)_j.
    \label{eq:position-map-compact}
\end{equation}
Here $C$ is determined by the centered position operators introduced in Eq.~\eqref{eq:x_centered} and \eqref{eq:trajectory_in_u}.

This distinction becomes especially relevant in the small-$c$ regime. As established in Eq.~\eqref{eq:smallc-residence} and illustrated in Fig.~\ref{fig:smallc-fixed-couplings-summary}(b), the fraction of each driver period spent near the pure-state boundary $|\mathbf B|=1$ approaches unity as $c\to0^+$ at fixed theory parameters. Whether this near-boundary residence is visible in physical position observables depends on the rank and orientation of $C$.  If $C$ is invertible rank-3 matrix, the three position expectation values determine $\mathbf B(t)$ uniquely, so no information about the two-level density matrix is lost. However, a general invertible $C$ distorts the Euclidean Bloch geometry. In particular, the stroboscopic position map is
\[
    \mathbf r(t+T_\kappa)
    =C Q_F^B C^{-1}\mathbf r(t),
\]
The direction $C\mathbf n_F^B$ remains invariant under this map, but it is not in general an axis of a Euclidean spatial rotation. 

\begin{figure}[H]
\centering
    \includegraphics[width=10.0 cm]{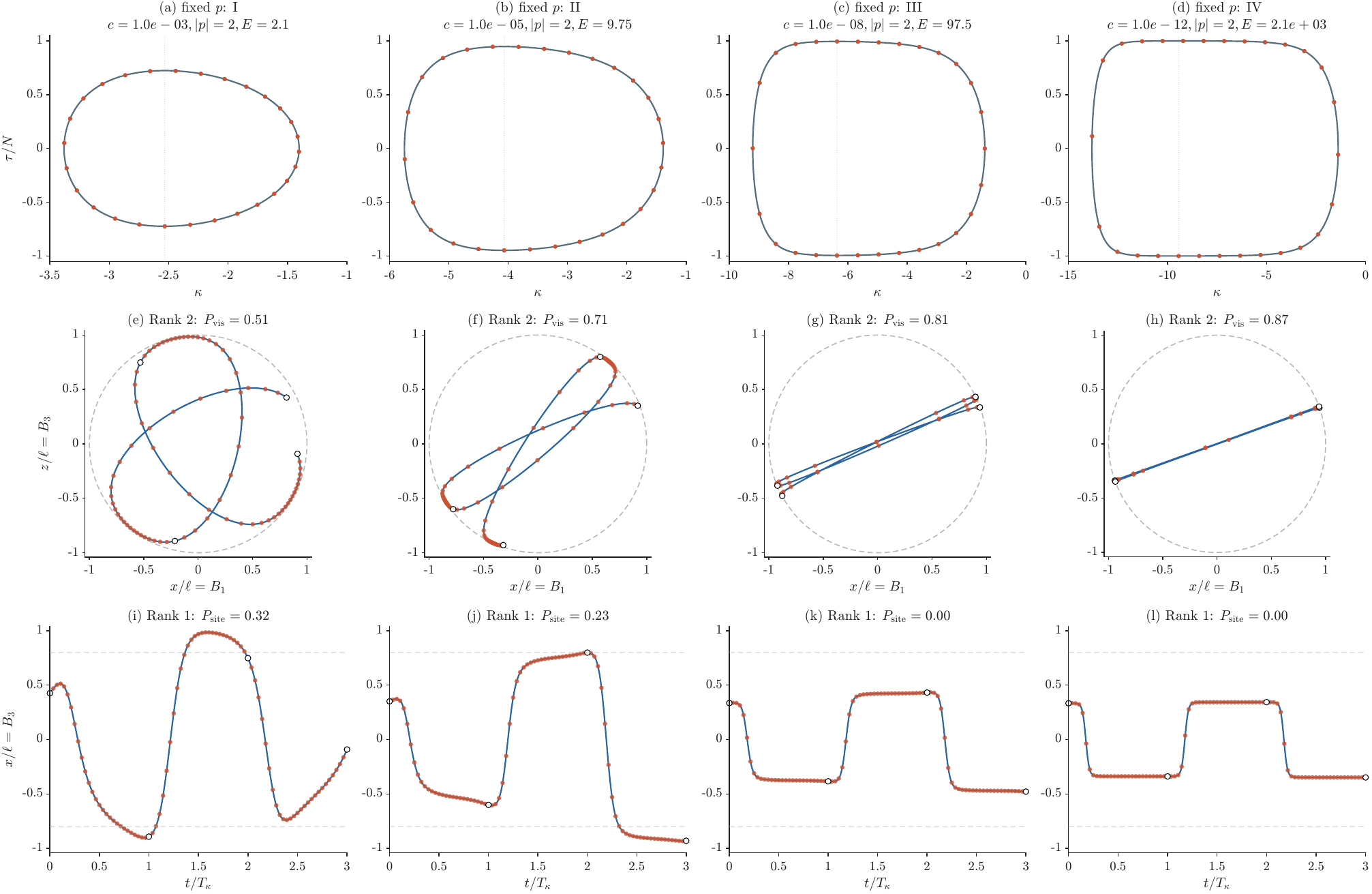}
    \caption{Physical position-expectation trajectories in the small-$c$ regime at fixed $p=-2$, $a=\lambda=0$, $b=N=1$. From left to right, $c=10^{-3},10^{-5},10^{-8},10^{-12}$, with the corresponding normalized energy $E$ increasing. The upper row shows the periodic $(\kappa,\tau/N)$ driver orbits. The middle row displays the rank-two position map $(x,z)/\ell=(B_1,B_3)$, while the lower row shows the rank-one position signal $x/\ell=B_3$. All trajectories cover three driver       periods; filled markers indicate equal physical-time intervals and open circles denote stroboscopic points $t=nT_\kappa$. The dashed circle represents the pure-state boundary in the rank-two plane, and the dashed horizontal lines mark $B_3=\pm0.8$ in the rank-one panels. The residence fractions are evaluated by uniform sampling in physical time over the displayed three-period interval: $P_{\mathrm{vis}} =P(\sqrt{B_1^2+B_3^2}\geq0.95)$ and $P_{\mathrm{site}}=P(|B_3|\geq0.8)$. Since $B_2=0$ in this sector, $P_{\mathrm{vis}}$ coincides with the Bloch-radius residence probability at $R_0=0.95$, discussed in Eq.~\eqref{eq:smallc-residence}. The rank-two trajectory therefore exhibits the predicted increasing residence near the pure-state boundary, whereas the rank-one readout need not concentrate near its measurement endpoints.
    }
    \label{fig:smallc-physical-rank12}
\end{figure}

In this section, we are interested in how the near-boundary small-$c$ dynamics may change when transitioning to physical space observables; therefore, we consider a simplified case, without the additional dynamic effects that arise when the parameter $a$ is included. With this choice, both state vectors are real, as do the Bloch vectors; consequently, the dynamics effectively become two-dimensional, and we can limit ourselves to considering rank-2 and rank-1 observables. Figure~\ref{fig:smallc-physical-rank12} illustrates rank-two and rank-one maps applied to the same underlying nonlinear dynamics.

\paragraph{Rank-two map: planar position trajectory.}

Consider the following position observables\footnote{Here the position observables are understood as the matrix representations of the physical position operators in the two-dimensional subspace spanned by $\{|A_1\rangle,|A_2\rangle\}$.}:
\begin{equation}
    X=\ell\sigma_1,
    \qquad
    Y=0,
    \qquad
    Z=\ell\sigma_3,
\end{equation}
which give
\begin{equation}
    \mathbf r(t)
    =
    \ell
    \begin{pmatrix}
        B_1(t)\\
        0\\
        B_3(t)
    \end{pmatrix}.
    \label{eq:rank2-map-compact}
\end{equation}
The position-expectation trajectory is, therefore, confined to the $(x,z)$ plane. In the real sector $a=\lambda=0$, $B_2(t)=0$, so this map retains the entire Bloch trajectory and its amplitude is governed directly by the Bloch radius:
\begin{equation}
    \frac{|\mathbf r(t)|}{\ell}
    =\sqrt{B_1^2(t)+B_3^2(t)}
    =|\mathbf B(t)|=R_B(t).
\end{equation}
Consequently, the Bloch-space residence probability derived earlier has a direct physical space interpretation: the fraction of time spent at $R_B>R_0$ equals the fraction spent at $|\mathbf r|>\ell R_0$.  As $c\to0^+$, the position-expectation trajectory spends an increasing fraction of each driver period near the circle $|\mathbf r|=\ell$. This behavior is visible in the middle row of Fig.~\ref{fig:smallc-physical-rank12}, where markers placed at equal physical-time intervals illustrate the increasing residence near the outer circle.

As $a=\lambda=0$ in the examples shown in Fig.~\ref{fig:smallc-physical-rank12}, the canonical Floquet axis is $\mathbf n_F^B=\pm(0,1,0)^T$, i.e. it is normal to the observed $(B_1,B_3)$ plane. Consequently, the stroboscopic Floquet map acts as a planar rotation of the position-expectation trajectory by $2\theta_F$ modulo $2\pi$, with the sign fixed by the axis orientation. The open markers in the middle row indicate successive positions after each driver period, illustrating how the observed trajectory is reoriented.

\paragraph{Rank-one map: one-dimensional position readout.}

For comparison, consider
\begin{equation}
    X=\ell\sigma_3,
    \qquad
    Y=Z=0,
\end{equation}
so that
\begin{equation}
    x(t)=\ell B_3(t).
    \label{eq:rank1-map-compact}
\end{equation}
This provides a one-dimensional position readout, which may represent a two-site or double-well system whose basis states are localized near $x=\pm\ell$. The expectation value measures the population imbalance, while the transverse Bloch components remain unobserved.

Unlike the rank-two case, proximity to the pure-state boundary does not automatically imply the proximity to the endpoints $x=\pm\ell$. Indeed, $|\mathbf B|\simeq1$ is compatible with $|B_3|\ll1$ when the transverse components carry most of the Bloch-vector norm. This difference is illustrated by the lower row of Fig.~\ref{fig:smallc-physical-rank12}. The indicated site-residence fraction $P_{\mathrm{site}}=P(|B_3|\geq0.8)$ measures the fraction of the displayed time interval for which the position expectation value lies near either endpoint. Unlike $P_{\mathrm{vis}}$, its behavior depends on the angular distribution of the Bloch trajectory. Consequently, increasing residence near $|\mathbf B|=1$ does not necessarily correspond to an analogous increase in residence near $x=\pm\ell$.

The canonical Floquet axis $\mathbf n_F^B$ continues to govern the underlying stroboscopic rotation, but the rank-one readout reveals only its projection onto the measured direction.

To conclude, the three rows of Fig.~\ref{fig:smallc-physical-rank12} provide complementary views of the same dynamics. The upper row shows the periodic nonlinear driver, the middle row retains the full Bloch trajectory in a planar position map, and the lower row displays its one-dimensional $B_3$ readout. All panels cover three driver periods, with filled markers at equal physical-time intervals and open markers indicating stroboscopic sampling. The two residence fractions probe different physical properties of the same dynamics. Since $B_2=0$ and $R_B(t)$ is periodic, $P_{\mathrm{vis}}$ coincides with the one-period Bloch-radius residence probability $\mathcal P_c(R_B>0.95)$, which approaches unity as $c\to0^+$. Meanwhile, $P_{\mathrm{site}}$ measures the fraction of time for which one of the two position-basis states has occupation probability at least $90\%$, corresponding to $|B_3|\geq0.8$. Such localization is not implied by near-purity, since a pure state may remain a coherent superposition of the two sites.

\section{Conclusions}
\label{sec:conclusions}
In the preceding sections, we have advanced our understanding of a previously introduced non-linear extension of quantum mechanics. As reviewed in the introduction and in section \ref{sec:model_review}, the non-linear aspects involve a pair of state vectors that describe a given system, in contrast to ordinary quantum mechanics in which a single state vector can encode the behavior of a pure state. In section \ref{sec:analytic_res}, we showed that when a certain integration constant, $c$, is set to zero, the system reverts to ordinary quantum mechanics. This is essentially because when $c=0$ the two state vectors are forced to be proportional. Whatever the nature of non-linear effects in quantum mechanics, they must be very small, so the $c\rightarrow0^+$ limit suggests itself as a regime worthy of further study. 

In \cite{ChodosCooper2025} a simple exact solution was studied in detail, and here we have leveraged that analysis by constructing a lowest -order perturbation around it. In section \ref{sec:numeric_res} we compare the perturbation solution to a numerical integration of the exact equations and find excellent agreement between them. 

Living within our model is an independent dynamical system: a one-dimensional classical particle with coordinate $\kappa(t)$ moving in an exponential potential. With suitable choice of parameters, the motion is periodic and bounded. The driver motion's function enters the equations for the state vectors themselves, which can be reduced to a pair of first-order evolution equations that are periodically driven at the frequency of the classical particle’s motion. We analyze these equations in section \ref{sec:floquet_structure} using Floquet theory and study the interplay between the periodic driver and the corresponding Floquet dynamics. 
Although the coefficient matrix of the linear subsystem is generally not anti-Hermitian, the imposed theory constraints imply that the one-period monodromy is actually similar to an $SU(2)$ matrix. The resulting evolution of the density matrix between successive driver periods is then described by a fixed unitary conjugation. In particular, the nonlinear driver modulates the Bloch radius during each period, while the Floquet eigenphase and a corresponding axis governs the angular evolution. The resulting structure also provides exact recurrence conditions for quadratic observables when the principal Floquet ratio is rational.

Because the dynamics of the driver is independent of the details of the underlying system, our model possesses a certain universality, which motivates us to think of the non-linear effects as possibly connected to gravitation. Of course, a successful classical theory of gravity, general relativity, already exists, and it has not been explicitly included in our model. There are, however, many examples of dualities, both in ordinary quantum mechanics and in more advanced settings such as string theory, that allow for complementary descriptions of the same phenomena. Given the difficulties with quantizing general relativity, it may be worthwhile to see if some aspects of gravity can be understood within the context of our model.

To pursue this idea, we assume the underlying system possesses a number of position operators (most simply, this could be just a free particle moving in D dimensions) and we compute the influence of the non-linear dynamics on the evolution of their expectation values. This exercise was carried out explicitly in \cite{ChodosCooper2025} for the simple solution studied there; here we extend this analysis numerically to some more complicated cases. 
Since the system is non-linear, the resulting position-expectation trajectories exhibit a variety of nontrivial spatial patterns, including petal-like and fan-like structures, whose geometry reflects both the underlying Floquet dynamics and the choice of position observables.
Furthermore, our analysis of the approach to ordinary quantum mechanics reveals a nonuniform limiting behavior. As $c\to0^+$ at fixed model parameters, the reduced density matrix remains nearly pure for an increasingly large fraction of each driver period, while still undergoing periodic excursions toward the maximally mixed state. These excursions occupy a vanishing fraction of the period, which itself diverges logarithmically in this limit. Consequently, the approach to the pure-state sector is not uniform over the full driver cycle. For suitably chosen position observables, the increasing residence near the pure-state boundary is also reflected in the geometry of the corresponding expectation-value trajectories. We further find that the limiting dynamics depends on the path taken through parameter space, particularly when the limit $c\to0^+$ is accompanied by an approach to the exceptional surface $b+\mu=0$.

However, the parameter space available for exploration is quite large: the equations of motion themselves contain four free parameters, and two more constants appear in the process of integrating them. We cannot conclude that the trajectories we have already examined are representative of the full range of possibilities. If our model has any application to the real world, the non-linear effects must be very small on the atomic or sub-atomic scale, which implies that the most relevant corners of parameter space are those (such as $c\rightarrow0$) that closely approach ordinary quantum mechanics. We intend to identify and study such regions in future work.

Finally, we comment further on a point raised in the introduction. As mentioned there, Weinberg’s model was effectively killed when Gisin and Polchinski showed that the measurement process could be used for superluminal communication. The prejudice against this possibility is strong. As expressed in one paper \cite{HelouChen2017}, "the condition…that one cannot send information faster than the speed of light…is a cornerstone of the special theory of relativity. The community regards this condition as being inviolable." But in physics, it is not the community that decides issues like these, it is experiment. 
Conventional wisdom \cite{Gisin1990, BassiHejazi2015} has it that to prevent superluminal signalling the evolution equation for the density operator $\hat \rho$ must depend linearly on $\hat \rho$, and not separately on the state vectors. In our model,
\begin{equation}
    \hat \rho = \frac{1}{N }\left[ |\psi_1\rangle \langle \psi_1| + |\psi_2\rangle \langle \psi_2| \right] 
\end{equation}
and its evolution is not reducible to a simple linear dependence \cite{ChodosCooper2025}
\begin{equation}
    i \frac{d \hat \rho}{dt} = \left[ H, \hat \rho \right] + \frac{2i\mu}{N} \left[ |\psi_1 \rangle \langle \psi_2 |, |\psi_2 \rangle \langle \psi_1 |\right] + \frac{2 \lambda}{N} \left[ |\psi_2 \rangle \langle \psi_2 |, |\psi_1 \rangle \langle \psi_1 | \right].
\end{equation}
This equation also shows that the chosen density matrix does not, by itself, provide a complete description of the system. Distinct pairs of state vectors can yield the same density operator but different time derivatives, so that specifying only $\hat\rho(0)$ is insufficient to determine its subsequent evolution. This dependence on the underlying state vectors, indeed, raises the possibility of superluminal communication. However, the matter is not completely settled, because the two state vectors do not themselves evolve independently, and one needs to specify exactly how the pair of state vectors collapses when a measurement is made. This will be the subject of further investigation.
\begin{acknowledgments}
AC is grateful for contributions made by Fred Cooper and John Dawson at an earlier stage of this project.
\end{acknowledgments}

%%%%%%%%%%%%%%%%%%%%%%%%%%%%%%%%%%%%%%%%%%
%% Optional
\appendix
\counterwithin*{equation}{section} % reset 'equation' counter whenever '\section' is executed
\renewcommand{\theequation}{\Alph{section}\arabic{equation}}

\section[\appendixname~\thesection]{First-order correction for the Floquet axis}
\label{app:floquet_axis_correction}
The first-order fundamental matrix over one small-oscillation period $T_0$ is
\begin{equation}
    \mathcal{Z}^{(1)}(T_0)\equiv Z_1(T_0)
    =
    e^{M_0T_0}
    \left[
        \mathbb{I}+R F(T_0)
    \right]
    +O(R^2)\quad \text{ with } \quad 
    F(t)
    =
    -\frac{i e^{\kappa_0/2}}{2}
    \left(
       J_1\vec k_1
       +
       J_2\vec k_2
       +
       J_3\vec k_3
    \right)
    \cdot\bm{\sigma}.
    \label{eq:F_axis}
\end{equation}
The vectors $\vec k_i$ are defined by:
\begin{align}
    \vec k_1
    =
    \vec b+
    \frac{\vec a\cdot\vec b}{\omega_1^2}\vec a,\quad
    \vec k_2 =
    \frac{i}{\omega_1}
    \left(\vec b\times\vec a\right),\quad
    \vec k_3=
    -\frac{\vec a\cdot\vec b}{\omega_1^2}\vec a.
\end{align}
For $\lambda=0$ the components of $\vec b$ are obtained in the section before:
\begin{align}
    b_1
    =
    c_1\cos\widetilde\theta_0
    -
    c_2\sin\widetilde\theta_0,\quad
    b_2
    =
    c_2\cos\widetilde\theta_0
    +
    c_1\sin\widetilde\theta_0,\quad
    b_3=0,
\end{align}
where
\begin{equation}
    c_1
    =
    a\left(1+\frac{b}{b+\mu}\right),
    \qquad
    c_2
    =
    \frac{a^2}{b+\mu}-b.
\end{equation}
Two useful identities then follow:
\begin{equation}
    \mathbf n_0\cdot\vec b=\sqrt{a^2+b^2},
    \qquad
    \mathbf m_0\cdot\vec b=\frac{a\sqrt{a^2+b^2}}{b+\mu}.
    \label{eq:b_projections}
\end{equation}
Evaluation of the $J$-integrals at one $\kappa$-period $T_0$ gives:
\begin{align}
    J_1(T_0)
    &=
    \frac{
       \omega\sin\theta_0(1-\cos4\pi \frac{\omega_1}{\omega})
       -
       2\omega_1\cos\theta_0\sin4\pi \frac{\omega_1}{\omega}
    }{
       \omega^2-4\omega_1^2
    },
    \label{eq:J1_period}\\[1mm]
    J_2(T_0)
    &=
    -
    \frac{
       2\omega_1\cos\theta_0(1-\cos4\pi \frac{\omega_1}{\omega})
       +
       \omega\sin\theta_0\sin4\pi \frac{\omega_1}{\omega}
    }{
       \omega^2-4\omega_1^2
    },
    \label{eq:J2_period}\\
    J_3(T_0)&=0.
    \label{eq:J3_period}
\end{align}
Transforming Eq.~\eqref{eq:F_axis} with $L_0$ that we have defined previously in Eq.~\eqref{eq:L0_axis}, one finds
\begin{align}
    \widetilde{\vec k}_1
    &=
    \frac{a\sqrt{a^2+b^2}}{b+\mu}\mathbf m_0
    -
    i \sqrt{a^2+b^2}\sinh\Lambda\,\mathbf e_3,\\
    \widetilde{\vec k}_2
    &=
    -\frac{a\sqrt{a^2+b^2}}{b+\mu}\mathbf e_3
    -
    i \sqrt{a^2+b^2}\sinh\Lambda\,\mathbf m_0,\\
    \widetilde{\vec k}_3
    &=
    \sqrt{a^2+b^2}\cosh\Lambda\,\mathbf n_0 .
\end{align}
To isolate the correction to the rotational Floquet dynamics, we take the
polar decomposition of the first-order factor. For some generic matrix
\(A =  \left( \mathbb{I}+RF \right)+O(R^2)
\), this decomposition can be written as follows:
\[
A=UP,\qquad
U^\dagger U=\mathbb{I},\qquad P=P^\dagger>0,
\]
with rotational and scaling parts defined accordingly:
\[
U
=
\mathbb{I}+\frac{R}{2}(F-F^\dagger)+O(R^2),
\qquad
P
=
\mathbb{I}+\frac{R}{2}(F+F^\dagger)+O(R^2).
\]
Thus, the anti-Hermitian part of $F$ determines the first-order correction
to the unitary rotational factor, whereas the Hermitian part contributes to
the positive scaling factor. In the present case, after the $L_0$ transform for the first order correction to the fundamental matrix we have:
\[
\widetilde F
=
-\frac{i e^{\kappa_0/2}}{2}
\frac{a\sqrt{a^2+b^2}}{b+\mu}
\left(
J_1\mathbf m_0-J_2\mathbf e_3
\right)\cdot\boldsymbol\sigma
-
\frac{e^{\kappa_0/2}}{2}
\sqrt{a^2+b^2}\sinh\Lambda
\left(
J_2\mathbf m_0+J_1\mathbf e_3
\right)\cdot\boldsymbol\sigma .
\]
The first term is anti-Hermitian and the second is Hermitian.  Hence, the relevant correction to the Floquet axis is defined by:
\begin{equation}
    \frac12
    \left(
        \widetilde F-\widetilde F^\dagger
    \right)
    =
    -i\,\mathbf d\cdot\boldsymbol{\sigma}, \text{ with }
    \mathbf d
    =
    \frac{ae^{\kappa_0/2}\sqrt{a^2+b^2}}{2(b+\mu)}
    \left(
        J_1\,\mathbf m_0
        -
        J_2\,\mathbf e_3
    \right).
    \label{eq:u_axis}
\end{equation}
Since both $\mathbf m_0$ and $\mathbf e_3$ are orthogonal to
$\mathbf n_0$, one has $\mathbf d\cdot\mathbf n_0=0$.  The first-order
unitary correction therefore tilts the effective Floquet rotation axis,
while the corresponding correction to the Floquet eigenphase vanishes at
this order away from a zone-edge degeneracy.

Consequently the rotational part of the one-period evolution is
\begin{equation}
    \mathcal{M}^{(1)}
    =
    \exp\left\{-i\omega_1 T_0~\mathbf n_0\cdot\bm{\sigma}\right\}
    \left[
        \mathbb{I}-iR~\mathbf d\cdot\boldsymbol{\sigma}
    \right]
    +O(R^2).
    \label{eq:UF_first}
\end{equation}
Away from the resonance zone, where $\sin \omega_1 T_0 \neq0$, we may write
\begin{equation}
    \mathbf n_F^{(\eta_0)}
    =
    \mathbf n_0
    +
    R\,\delta\mathbf n_F^{(1, \eta_0)}
    +O(R^2).
\end{equation}
Expanding Eq.~\eqref{eq:UF_first} with the Pauli identity
\begin{equation}
    (\mathbf x\cdot\boldsymbol{\sigma})
    (\mathbf y\cdot\boldsymbol{\sigma})
    =
    (\mathbf x\cdot\mathbf y)I
    +
    i(\mathbf x\times\mathbf y)
       \cdot\boldsymbol{\sigma}, \nonumber
\end{equation}
gives
\begin{equation}
    \delta\mathbf n_F^{(1, \eta_0)}
    =
    \mathbf n_0\times\mathbf d
    +
    \cot (\omega_1 T_0)
    \left[
        \mathbf d
        -
        (\mathbf n_0\cdot\mathbf d)\mathbf n_0
    \right]
    =
    \mathbf n_0\times\mathbf d
    +
    \cot (\omega_1 T_0)\,\mathbf d.
\end{equation}
The one-period integrals satisfy the identities
\begin{align}
    J_2+J_1\cot (\omega_1 T_0)
    =
    -\frac{4\omega_1\cos\theta_0}
          {\omega^2-4\omega_1^2},\quad
    J_1-J_2\cot (\omega_1 T_0)
    =
    \frac{2\omega\sin\theta_0}
         {\omega^2-4\omega_1^2},
\end{align}
which finally yield
\begin{equation}
    \delta\mathbf n_F^{(1,\eta_0)}
    =
    \frac{ae^{\kappa_0/2}\sqrt{a^2+b^2}}
         {(b+\mu)\left(\omega^2-4\omega_1^2\right)}
    \left[
       -2\omega_1\cos\theta_0\,\mathbf m_0
       +
       \omega\sin\theta_0\,\mathbf e_3
    \right].
    \label{eq:dn_final}
\end{equation}
Thus the first finite-amplitude correction has a particularly simple
geometrical meaning: it consists of an azimuthal rotation in the equatorial
plane, along $\mathbf m_0$, and a polar tilt along $\mathbf e_3$.

\section[\appendixname~\thesection]{What is, and what is not, Hermitian}
\label{sec:app_B}
The defining equations of motion are
\begin{equation}
    i \frac{d|\psi_1\rangle}{dt} = \left[ H + i\mu \langle \psi_2|\psi_2 \rangle \right] |\psi_1 \rangle + \left( g+\lambda \right) \langle \psi_2 | \psi_1 \rangle |\psi_2 \rangle
\end{equation}
and
\begin{equation}
    i \frac{d|\psi_2\rangle}{dt} = \left[ H - i\mu \langle \psi_1|\psi_1 \rangle \right] |\psi_2 \rangle + \left( g^*-\lambda \right) \langle \psi_1 | \psi_2 \rangle |\psi_1 \rangle.
\end{equation}
Letting
\begin{equation}
    | \Psi \rangle = \begin{pmatrix}
        |\psi_1 \rangle \\ |\psi_2 \rangle
    \end{pmatrix},
\end{equation}
we can express them in matrix form
\begin{equation}
    i \frac{d |\Psi \rangle}{dt} = H |\Psi \rangle + M |\Psi \rangle,
\end{equation}
where
\begin{equation}
    M = \begin{pmatrix}
        i \mu \langle \psi_2|\psi_2\rangle & (g+\lambda) \langle\psi_2| \psi_1 \rangle \\ (g^* - \lambda) \langle \psi_1 | \psi_2 \rangle & - i \mu \langle \psi_1| \psi_1\rangle
    \end{pmatrix}.
\end{equation}
Here the elements of $M$ are complex numbers, and the terms proportional to $\mu$ and $\lambda$ are anti-Hermitian whereas those proportional to $a$ and $b$ are Hermitian, where $g=a+ib$.

However, we can also write these equations as
\begin{equation}
    i \frac{d |\Psi \rangle}{dt} = H |\Psi \rangle + M' |\Psi \rangle,
\end{equation}
where the elements of $M'$ are operators in Hilbert space:
\begin{equation}
    M' = \begin{pmatrix}
        (g+\lambda)  |\psi_2\rangle \langle \psi_2| & i\mu |\psi_1\rangle \langle \psi_2| \\ -i \mu |\psi_2\rangle \langle \psi_1| & (g^*-\lambda) |\psi_1\rangle \langle \psi_1|
    \end{pmatrix}.
\end{equation}
The equations are the same, but now the terms proportional to $\mu$ and $\lambda$ are Hermitian, whereas those proportional to $b$ are anti-Hermitian. The terms proportional to $a$ are Hermitian in both representations.

The fact that there are Hermitian ways of representing all the terms may help explain the conservation of the norm $\langle \Psi | \Psi \rangle$, even though we do not have a single representation in which they are all simultaneously Hermitian. 

 \bibliography{nonlinear_quantum_references.bib}

\end{document}